\documentclass[pra,aps,nofootinbib,showpacs,floatfix,superscriptaddress,amsmath,twocolumn]{revtex4-1}

\usepackage{graphicx}
\usepackage{amsmath}
\usepackage{amssymb}
\usepackage{color}
\usepackage{bm}
\usepackage{hyperref}
\hypersetup{colorlinks,allcolors=black}
\usepackage{url}

\newcommand {\fabs}[1] {\left| #1 \right|}
\newcommand {\fnorm}[1] {\left\Vert #1 \right\Vert}

\newcommand {\fabsq}[1] {\left\vert #1 \right\vert^2}

\newcommand{\ket}[1]{\ensuremath{|#1\rangle}}
\newcommand{\bra}[1]{\langle#1|}

\newcommand{\braket}[2]{\langle#1|#2\rangle}
\newcommand{\ketbra}[2]{|#1\rangle\langle#2|}
\newcommand{\braXket}[3]{\langle#1|#2|#3\rangle}

\newcommand{\Sket}[1]{\ensuremath{|#1\rangle\!\rangle}}
\newcommand{\Sbra}[1]{\langle\!\langle#1|}
\newcommand{\Sbraket}[2]{\langle\!\langle#1|#2\rangle\!\rangle}

\newcommand{\cB}{{\cal{B}}}
\newcommand{\cC}{{\cal{C}}}
\newcommand{\cE}{{\cal{E}}}
\newcommand{\cL}{{\cal{L}}}
\newcommand{\cH}{{\cal{H}}}

\newcommand{\cO}{{\cal{O}}}

\newcommand{\Eqref}[1]{Eq. \eqref{#1}}

\newcommand{\Hamil}{\mathcal{H}}
\newcommand{\fid}{\text{F}}

\newcommand\imageScale{0.9}

\begin{document}
\title{Robustness of Enhanced Shortcuts to Adiabaticity in Lattice Transport}
\author{C. Whitty}
\email{c.whitty@umail.ucc.ie}
\affiliation{Department of Physics, University College Cork, Cork, Ireland}

\author{A. Kiely}
\affiliation{School of Physics, University College Dublin, Belfield, Dublin 4, Ireland}
\affiliation{Centre for Quantum Engineering, Science, and Technology, University College Dublin, Belfield, Dublin 4, Ireland}

\author{A. Ruschhaupt}
\affiliation{Department of Physics, University College Cork, Cork, Ireland}

\begin{abstract}
Shortcuts to adiabaticity (STA) are a collection of quantum control techniques that achieve high fidelity outside of the adiabatic regime.
Recently an extension to shortcuts to adiabaticity was proposed by the authors [Phys. Rev. Research 2, 023360 (2020)].
This new method, enhanced shortcuts to adiabaticity (eSTA), provides an extension to the original STA control functions and allows effective control of systems not amenable to STA methods.
It is conjectured that eSTA schemes also enjoy an improved stability over their STA counterparts. We provide numerical evidence of this claim by applying eSTA to fast atomic transport using an optical lattice, and evaluating appropriate stability measures. 
We show that the eSTA schemes not only produce higher fidelities, but also remain more stable against errors than the original STA schemes.
\end{abstract}
\maketitle

\section{Introduction}

Quantum devices and technologies have the potential to revolutionize a broad range of scientific and engineering disciplines \cite{acinQuantumTechnologiesRoadmap2018,schleichQuantumTechnologyResearch2016a}.
However, fast and stable control of these systems is a significant barrier to building practical devices \cite{vanfrankOptimalControlComplex2016a}.
Quantum control needs to be fast to avoid decoherence, while simultaneously being robust and stable against implementation errors.
Furthermore, it should be effective within the constrained resources of the physical implementation, such as energetic cost or pulse bandwidth \cite{waltherControllingFastTransport2012}.

Often practical quantum control relies heavily on numerical optimization, which has some drawbacks: it can be computationally expensive, difficult to scale and it may be unclear how to understand or generalize the resulting schemes \cite{abdelhafezGradientbasedOptimalControl2019a,kellyOptimalQuantumControl2014,machnesTunableFlexibleEfficient2018}.
There exist alternative analytic techniques, such as Shortcuts to Adiabaticity (STA), which give exact quantum state transfer.
However, analytic techniques such as STA are known exactly for only a limited number of physical systems \cite{chenFastOptimalFrictionless2010b,torronteguiChapterShortcutsAdiabaticity2013a,guery-odelinShortcutsAdiabaticityConcepts2019b}. 
Recently an extension to STA methods has been developed, known as enhanced Shortcuts to Adiabaticity (eSTA), that provides control of a broader class of systems \cite{whittyQuantumControlEnhanced2020}.
Crucially, eSTA is an analytic technique that allows physical insight into the control scheme.
Additionally, eSTA has a much lower computation cost compared with full numerical optimization.
eSTA has been applied to population inversion in a two-level system without the rotating-wave approximation and to the transport of a Gaussian trap with one and two ions \cite{whittyQuantumControlEnhanced2020}.
Recently, there has already been a first application of eSTA techniques to the transport of atoms in an optical lattice \cite{hauck2021singleatom}.

STA schemes are extremely robust to noise and systematic errors \cite{ruschhauptOptimallyRobustShortcuts2012a,kielyFastStableManipulation2015b,luFastShuttlingTrapped2014,nessRealisticShortcutsAdiabaticity2018a,qiFastRobustParticle2021}.
An obvious question is whether robustness remains in the eSTA framework.
Therefore, the main goal of this paper is to provide a thorough investigation into how this stability is affected by the application of eSTA.

We choose atomic transport using an optical lattice to examine the robustness of eSTA, since the coherent control of lattice systems has many practical quantum technological applications.
For example: transport in atomic chains \cite{gutierrez-jaureguiCoherentControlAtomic2021}, modeling condensed matter systems \cite{karamlouQuantumTransportLocalization2021}, many-body phenomena in ultracold gases \cite{blochManybodyPhysicsUltracold2008} and the trapping and control of ions \cite{schneiderOpticalTrappingIon2010,linnetPinningIonIntracavity2012}.
Recently, there has been experimental exploration of atom transport via a lattice potential around the quantum speed limit \cite{peterDemonstrationQuantumBrachistochrones2021}.
We consider similar physical parameters with transport times near this proposed speed limit \cite{peterDemonstrationQuantumBrachistochrones2021}.
We then show that the eSTA schemes provide improved robustness over the corresponding STA schemes, when considering a variety of imperfections.

In the following section we give a brief review of the formalism of eSTA.
In Sec. \ref{sect_physical_setting} we present the physical optical lattice model we are considering; we will introduce different control schemes and we examine and compare the fidelities achieved by eSTA.
In Sec. \ref{sect_C_Q}, we first define and then examine the deviation of the eSTA control function under variations of the Hamiltonian.
This gives a preliminary indication of the stability of eSTA.
In the following sections we compare the stability of eSTA and STA in more detail.
This includes a definition of a sensitivity quantity and an error bound for a quantitative comparison of the stability of different schemes.
In Sec. \ref{sect_robustness_systematic} we consider systematic errors during the transport.
In Sec.  \ref{sect_robustness_noise} the stability of lattice transport is examined for noisy fluctuations.

\section{eSTA Formalism \label{sect_review}}

The purpose of eSTA is to provide a formalism by which existing STA methods can be extended to quantum control problems beyond their current scope. 

We start with a system with Hamiltonian $\cH_s$ that has a difficult quantum control problem.
However, we assume that the Hamiltonian $\cH_s$ can be approximated by another  Hamiltonian $\cH_0$ of an idealized system that has an exact STA solution.
In the following we construct the improved eSTA protocol for  $\cH_s$ using the solutions of the idealized system $\cH_0$.

\subsection{Construction of eSTA control scheme}

To make the approximation of $\cH_s$ by $\cH_0$ precise, we assume there is a parameter $\mu$ and Hamiltonians $\cH_\mu$ such that $\cH_{\mu=\mu_s} =\cH_s$ and $\cH_{\mu = 0} =\cH_0$.
We also assume we can parameterize the control scheme by a vector $\vec{\lambda}$.
We set $\vec{\lambda}_0$ to denote the STA scheme, and $\vec{\lambda}_s$ as the eSTA scheme.

\begin{figure}[t]
\begin{center}
(a)\includegraphics[width=0.44\linewidth]{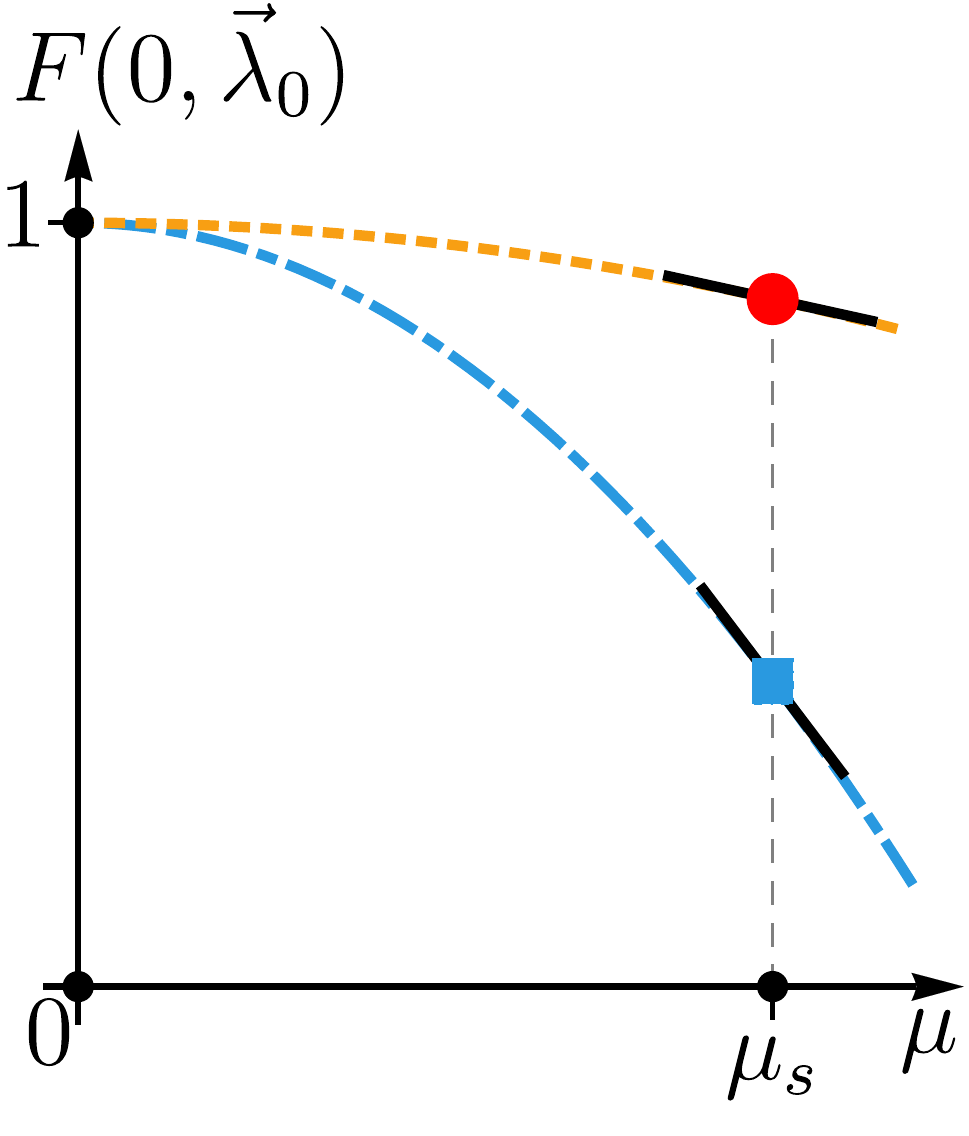}
(b)\includegraphics[width=0.44\linewidth]{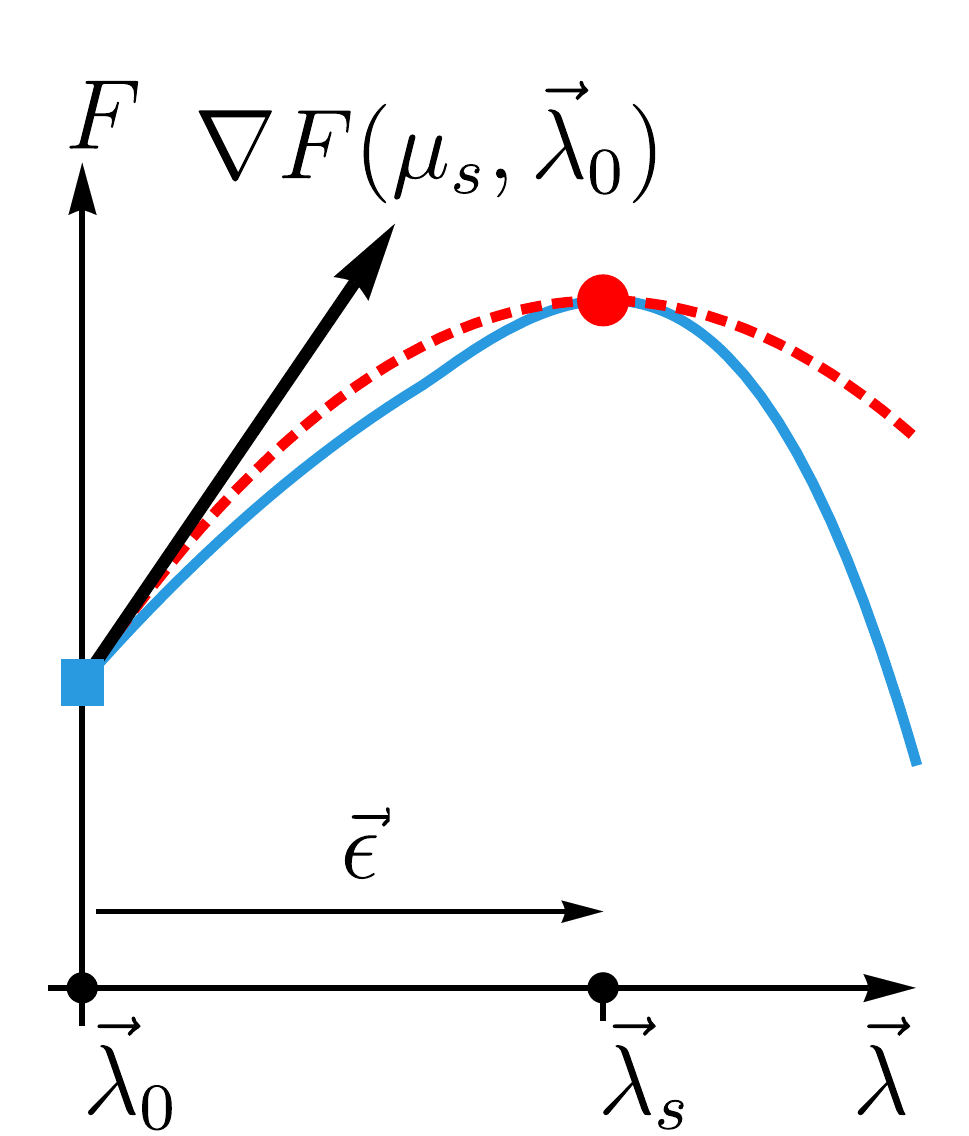}
\end{center}
\caption{\label{fig_1_schematic}(color online)
(a) is a schematic representation of fidelity versus $\mu$.
The fidelity of $\cH_\mu$ using $\vec{\lambda}_0$ is the dot-dashed blue line.
The blue square is $\cH_{\mu_s}$ using $\vec{\lambda}_0$.
The red dot is the improved fidelity of $\cH_{\mu_s}$ using $\vec{\lambda}_s$.
The dashed-orange line represents the assumed parabolic profile of the fidelity as $\cH_{0} \rightarrow \cH_{\mu_s}$, with the improved eSTA control $\vec{\lambda}_{\mu}$ calculated for each $\mu$.
The slopes of each line at $\mu_s$ are depicted as solid-black lines.
(b) is a diagram of eSTA in control space $(\vec{\lambda},F)$.
The starting STA scheme fidelity (blue square), gradient (black arrow) and fidelity at $\vec{\lambda}_s$ (red dot) are shown.
The resulting eSTA approximate parabola (dashed-red line), and true fidelity landscape (solid-blue line) are also displayed.
}
\end{figure}

Our goal is to evolve the initial state $\ket{ \Psi_0}$ at time $t=0$ to the target state $\ket{ \Psi_T}$ in a given total time $t_f$.
As previously discussed, we assume there exists an idealized system with known STA solutions and Hamiltonian $ \Hamil_0 (\vec{\lambda}_0; t)$.
$\vec{\lambda}_0$ solves the control problem for $\Hamil_0$, and we assume it works approximately for $\Hamil_{\mu_s}$.
To state this formally we define the fidelity
\begin{eqnarray}
\text{F}(\mu,\vec{\lambda})=
\fabsq{ \bra{\Psi_T}
	U_{\mu, \vec \lambda}(t_f,0) \ket{\Psi_0} },
\end{eqnarray}
where $U_{\mu, \vec \lambda}$ is the time evolution operator using the Hamiltonian $\Hamil_{\mu}$ with control scheme $\vec \lambda$.
We have that $\text{F}(\mu_s,\vec{\lambda}_0) < \text{F}(0,\vec{\lambda}_0) = 1$; see also Fig. \ref{fig_1_schematic} (a).

The goal of eSTA is to produce a $\vec{\lambda}_s$ that is built upon $\vec{\lambda}_0$, such that we maximize $\text{F}(\mu_s,\vec{\lambda_s})$ with $\text{F}(\mu_s,\vec{\lambda_s}) > \text{F}(\mu_s,\vec{\lambda_0})$.
We define $\vec{\lambda}_s = \vec{\lambda}_0 + \vec{\epsilon}$, and eSTA is now used to calculate $\vec{\epsilon}$.

To find $\vec{\epsilon}$, we use information about the fidelity and the gradient of the fidelity with respect to $\vec{\lambda}$ to construct a parabola in the parameter space of $\vec{\lambda}$ and $F$, as illustrated in Fig. \ref{fig_1_schematic} (b).
We assume that this parabola is a good approximation to the fidelity landscape in the direction of the gradient at $(\mu_s,\vec{\lambda_0})$.
This is schematically shown in Fig. \ref{fig_1_schematic} (b), with the dashed red line representing the parabolic approximation and the solid-blue line denoting the actual fidelity landscape.
Furthermore we assume that the maximum of this parabola gives perfect fidelity, i.e. $F(\mu_s,\vec{\lambda}_0+\vec{\epsilon})\approx 1$.
We can then write
\begin{eqnarray}
\vec\epsilon \approx 
\frac{2\left[1-\text{F}(\mu_s,\vec \lambda_0)\right]}{\left|\nabla_{\vec \lambda} F(\mu_s,\vec \lambda_0)\right|}
\frac{\nabla_{\vec \lambda} F(\mu_s,\vec \lambda_0)}{\left|\nabla_{\vec \lambda} F(\mu_s,\vec \lambda_0)\right|}. \label{parab}
\end{eqnarray}

We now calculate $\vec{\epsilon}$ using an approximation of the gradient and fidelity at the point $(\mu_s,\vec{\lambda}_0)$, shown as the blue square in Fig. \ref{fig_1_schematic}.
We begin the derivation of these estimates by assuming that the initial state $\ket{\Psi_0}$ and the final target state $\ket{\Psi_T}$ are both the same for the $\cH_0$ and $\cH_{\mu_s}$ systems.
We assume that the idealized $\cH_0$ system can be treated with STA techniques, i.e. there is a solution $\ket{\chi_0 (t)}$ of the time evolution
leading to fidelity one, i.e. $\ket{\chi_0 (0)} = \ket{\Psi_0}$ and $\ket{\chi_0 (t_f)} = \ket{\Psi_T}$.
In addition, we assume that there are solutions of the time evolution of $\cH_0$ labeled $\{\ket{\chi_n (t)}\}_{n \in \mathbb{N}}$ such that $\{\ket{\chi_n (t)}\}_{n \in \mathbb{N}_0}$
form an orthonormal basis for solutions of the $\cH_{\mu_s}$ system.
So we have
\begin{align}
\ket{\chi_n (t)} = U_{0,\vec \lambda_0} (t,0) \ket{\chi_n (0)},\\
U_{0,\vec \lambda_0} (t,s) =\sum_{n} \ketbra{\chi_n(t)}{\chi_n(s)}.
\end{align}
Note that in the following examples, we will use invariant-based inverse engineering to design the STA solutions and in these cases
$\{\ket{\chi_n (t)}\}_{n \in \mathbb{N}_0}$ are (up to a phase) the instantaneous eigenstates of the corresponding invariant.

We use time dependent perturbation theory to calculate an approximation of the fidelity $F(\mu_s,\vec{\lambda}_0)$ and the gradient of the fidelity $\nabla F(\mu_s,\vec{\lambda}_0)$.
We assume that we can neglect higher order contributions in both $\mu$ and $\vec\epsilon$.
The details of the calculations can be found in \cite{whittyQuantumControlEnhanced2020}.
For the fidelity $F(\mu_s, \vec \lambda_0)$ we then obtain up to second order in $\mu_s$
\begin{align}
F(\mu_s, \vec \lambda_0)
&\approx 1-\frac{1}{\hbar^2} \sum_{n=1}^\infty \fabsq{G_n},
\end{align}
where
\begin{align}
G_n = \!\! \int_0^{t_f} \!\! dt \braXket{\chi_n(t)}{\left[\Hamil_{\mu_s} (\vec\lambda_0;t) - H_{0} (\vec\lambda_0;t)\right]}{\chi_0(t)}.
\label{eq:Gn}
\end{align}
For the gradient approximation we find
\begin{align}
\nabla F (\mu_s, \lambda_0)
&\approx
-\frac{2}{\hbar^2} \sum_{n=1}^\infty \text{Re} \left( G_n \vec{K}_n^* \right),
\end{align}
with
\begin{align}
\vec{K}_n 
&=
\int_0^{t_f} dt \, \bra{\chi_n (t)} \nabla \Hamil_{\mu_s} (\vec\lambda_0; t) \ket{\chi_0 (t)}.
\label{eq:Kn}
\end{align}

Using \Eqref{parab}, we can now write the analytical expression for the eSTA protocol $\vec\lambda_s \approx \vec\lambda_0 + \vec{\epsilon}$
\begin{align}\label{eq:epsilon}
\vec{\epsilon} = -\frac{\left(\sum_{n=1}^N \fabsq{G_n}\right)\left[\sum_{n=1}^N \mbox{Re} \left(G_n^* \vec{K}_n\right)\right]}
{\fabsq{\sum_{n=1}^N \mbox{Re} \left(G_n^* \vec{K}_n\right)}},
\end{align}
where $G_n$ is given by \Eqref{eq:Gn}, $K_n$ is given by \Eqref{eq:Kn}, and we have truncated the infinite sums to the first $N$ terms.

It is important to note that $G_n$ and $\vec{K}_n$ can both be easily calculated as only the Hamiltonians and the known STA solutions for the idealized system with Hamiltonian $\Hamil_0$ are needed.
We also note that eSTA can produce protocols which are outside the class of STA schemes and offers improvement over previous perturbation based optimization \cite{whittyQuantumControlEnhanced2020}.

\subsection{Expected stabilities of eSTA and STA}

We now provide a general heuristic argument of why it is expected that eSTA protocols not only have improved fidelity, but also have improved stability compared with their corresponding STA schemes.
Let us consider Fig. \ref{fig_1_schematic} (a), showing schematically the fidelity versus $\mu$; $\mu=0$ corresponds with the approximated system and $\mu_s$ the system of interest.
The fidelity of $\cH_\mu$ using the initial STA control scheme $\vec{\lambda}_0$ is the dot-dashed blue line; the dashed-orange line represents
the fidelity using the improved eSTA control $\vec{\lambda}_{\mu}$ calculated for each $\mu$.
Since the STA and eSTA control schemes agree at $\mu= 0$, the fidelity must be one for both at this point.
By construction, for a given $\mu$ the fidelity produced using the eSTA control scheme (dashed orange line) is higher than the corresponding one for the STA scheme (dot-dashed blue line).
Assuming that the fidelity behaves as a parabola, the magnitude of the slope of the eSTA fidelity must be less than the slope of the STA fidelity at $\mu_s$.

We now consider the effect of a systematic error $\delta$, such that the STA and eSTA schemes derived using $\mu_s$ are now applied to the
system at a different, near-by $\mu = \mu_s (1 + \delta)$. We expect that the derivative of the fidelity in both cases can be approximated
by the slopes shown in Fig. \ref{fig_1_schematic} (a) (solid-black lines). Therefore, the fidelity in the eSTA case should vary less than the fidelity in the STA case.
Thus, the eSTA protocol should have higher stability against changes in $\mu$ than the corresponding STA scheme.
To confirm this intuitive reasoning, in the following we will examine the stability in more detail.

\section{Physical Model \label{sect_physical_setting}}

We consider atomic transport in an optical lattice, which currently has no STA solution.
For sufficient trapping depths the lattice potential can be well approximated locally by the harmonic potential.
The transport of a harmonic potential has an exact STA solution for all transport times using Lewis-Riesenfeld invariants \cite{torronteguiChapterShortcutsAdiabaticity2013a,guery-odelinShortcutsAdiabaticityConcepts2019b}.
Hence, we choose the harmonic potential transport STA solutions as the starting point to produce an eSTA protocol for the lattice transport problem.

The lattice potential is given by
\begin{align}\label{eq:lattice_pot}
V_S (x) = U_0
\operatorname{sin}^2\left( k_0 x\right),
\end{align}
where $U_0= \alpha \text{E}_{\text{rec}}$, $\text{E}_{\text{rec}} = 2(\pi \hbar)^2 / m \lambda^2$ and $k_0 =2 \pi/\lambda$.
We choose $\alpha=150$ in agreement with a physically implemented value in \cite{peterDemonstrationQuantumBrachistochrones2021}.
The motion of the lattice is described by a function $q_0(t)$, and Hamiltonian $\cH_{\mu_s}=H_S$ where
\begin{align}\label{eq:H_s}
H_S = \frac{p^2}{2 m} + V_{S} \left[x-q_0(\vec{\lambda},t)\right].
\end{align}

For designing the STA trajectories, we apply a harmonic approximation to the potential $V_S$.
A series expansion of $V_S (x)$ results in
\begin{align}
V_S (x) = V_0 (x) 
+ \mathcal{O} \left(  x^{4} \right),
\end{align}
where
\begin{align}
V_{0} (x) &= \frac{1}{2} m \omega_0^2 x^2,
\end{align}
and $\omega_0 = \sqrt{\frac{2 U_0}{m}} k_0 = \sqrt{\alpha}(4 \pi^2 \hbar) / m \lambda^2$. 
The corresponding idealized Hamiltonian $\Hamil_{0}=H_0$ is
\begin{align}\label{eq:H_0}
H_{0} = \frac{p^2}{2 m} + V_{0} [x-q_0(\vec{\lambda},t)].
\end{align}
Note that in this case as $U_0 \rightarrow \infty$, $H_S \rightarrow H_0$.
 
We also define a time unit $\tau = 2 \pi /\omega_0$ and a spatial unit $\sigma = \sqrt{\hbar / (m \omega_0)}$.

\subsection{STA control functions}

There exists known STA techniques to design trajectories $q_0(t)$ that give fidelity $F=1$ for the harmonic potential $H_{0}$ and arbitrary transport times \cite{torronteguiChapterShortcutsAdiabaticity2013a,guery-odelinShortcutsAdiabaticityConcepts2019b}.
In the following, we will use Lewis-Riesenfeld invariants to obtain STA trajectories for $H_{0}$ \cite{torronteguiFastAtomicTransport2011,guery-odelinShortcutsAdiabaticityConcepts2019b}.
For harmonic trap transport a known dynamical invariant has the form \cite{guery-odelinShortcutsAdiabaticityConcepts2019b}
\begin{align}
I(t)=\frac{1}{2m}\left( p - m \dot{q}_c \right)^2
+\frac{1}{2}m\omega_0^2 \left[ x - m q_c\left(t\right) \right]^2,
\end{align}
where $q_c(t)$ must satisfy the auxiliary equation
\begin{align}\label{eq:aux}
\ddot{q}_c+\omega_0^2\left(q_c - q_0 \right) = 0.
\end{align}
Note that \Eqref{eq:aux} describes a single-particle classical equation of motion, where $q_0(t)$ is the trajectory of the potential minimum and $q_c(t)$ is the resulting classical particle trajectory.

Solutions of the Schr\"odinger equation $i \hbar \partial/\partial t \Psi(x,t) = H_0 \Psi(x,t)$ can be expressed in terms of weighted transport modes,
\begin{align}\label{eq:Schr_solus}
\Psi(x,t) = \sum_n c_n e^{i \theta_n(t)} \psi_n(x,t),
\end{align}
where $\psi_n(x,t)$ are orthonormal eigenstates of the invariant $I$ satisfying $I(t) \psi_n(x,t)=\lambda_n \psi_n(x,t)$, $c_n$ are constants and the Lewis-Riesenfeld phase is given by
\begin{align}
\theta_n(t) = \frac{1}{\hbar}
\int_0^t
\braXket{\phi(t',n)}{i\hbar \frac{\partial}{\partial t'}-H_0(t')}{\phi(t',n)} dt'.
\end{align}

In the specific case of harmonic transport, the resulting transport modes in \Eqref{eq:Schr_solus} are
\begin{align}\label{eq:transport_modes}
\chi_n(x,t)=
e^{i \theta_n(t)} \psi_n (x,t) = e^{i \theta_n(t)} e^{\frac{i}{\hbar} m \dot{q_c} x} \phi_n(x-q_c),
\end{align}
where
\begin{align}
\theta_n(t) = -\frac{i}{\hbar} [(n+1/2) \hbar \omega_0] t + \int_0^t \frac{m \dot{q_c}^2}{2}dt',
\end{align}
with $\lambda_n = (n+1/2) \hbar \omega_0$ and $\phi_n(x)$ are solutions to the Schr\"odinger equation at $t=0$, i.e. harmonic eigenstates.

To ensure $I(t)$ and $H_0(t)$ agree at initial and final times, we set $[I(t),H_0(t)]=0$ for $t=0,t_f$.
This is equivalent [via \Eqref{eq:aux} and \Eqref{eq:transport_modes}] to the boundary conditions
\begin{align}\label{eq:qc_boundary}
q_c(0) &= 0, &q_c(t_f) &= d, \nonumber \\
\dot{q}_c(0) &= \ddot{q}_c(0) =0, &\dot{q}_c(t_f) &= \ddot{q}_c(t_f) = 0.
\end{align}
The key idea is that $q_c(t)$ can be chosen first, for example to be a polynomial that satisfies the boundary conditions in \Eqref{eq:qc_boundary}, and then $q_0(t)$ can be inverse engineered using \Eqref{eq:aux}.

Throughout this paper we enforce further boundary conditions on $q_c(t)$, namely
\begin{align}\label{eq:qc_boundary2}
q_c^{(3)}(t') = q_c^{(4)}(t') = 0, \text{ for } t = 0,t_f,
\end{align}
so that the resulting trap trajectory $q_0(t)$ ensures the trap is at rest for initial and final times.
In the following, we will use three different auxiliary functions $q_{c}(t)$ and calculate the corresponding STA trajectories $q_{0}(t)$.

{\bf Polynomial Function $\bm{q_{c,1}(t)}$:}
One of the simplest choices for an auxiliary function is a polynomial ansatz $q_{c,1}(t)$ \cite{guery-odelinShortcutsAdiabaticityConcepts2019b}, i.e.
\begin{eqnarray}
q_{c,1}(t)=\sum_{j=1}^J a_j t^j
\end{eqnarray}
where $J$ is the number of boundary conditions.
For the boundary conditions in \Eqref{eq:qc_boundary} and \Eqref{eq:qc_boundary2}, $J=10$ and we solve for the $a_j$ to get $q_{c,1}(t)$.
We then use \Eqref{eq:aux} to produce $q_{0,1}(t)$.
Examples of $q_{c,1}(t)$ and $q_{0,1}(t)$ can be seen in Fig. \ref{fig_2_qc_and_q0} (a) and (b) respectively (blue dot-dashed lines).

{\bf Quasi-optimal function $\bm{q_{c,2}(t)}$:}
Moving beyond the simple polynomial ansatz for $q_c$, we consider a trajectory introduced in \cite{zhangOptimalShortcutsAtomic2016} as a quasi-optimal solution to minimizing the quartic term in the potential $(1/2) m \omega_0^2[x-q_0(t)]^2-\beta[x-q_0(t)]^4$.
We label this auxiliary function $q_{c,2}(t)$ and our motivation for using this function is that it should reduce the effect of the anharmonic contribution within $H_{s}$.
This auxiliary function was derived using Pontryagin's maximal principle, and relies on first calculating a function $f_c$ that minimizes the quartic term contribution to the potential during transport \cite{zhangOptimalShortcutsAtomic2016},
\begin{align}
f_c(t)&= \frac{3d}{8} \left( 1-2 \frac{t}{t_f} \right)^{7/3} + \frac{7d}{4}\frac{t}{t_f} - \frac{3 d}{8}.
\end{align}
This function $f_c$ does not satisfy the boundary conditions in \Eqref{eq:qc_boundary} and it is the root of a complex valued equation \cite{zhangOptimalShortcutsAtomic2016}.  
We simplify the definition of $f_c$, by mapping $f_c$ from $(0,t_f/2)$ to $(t_f/2,t_f)$ appropriately.
We also enforce the boundary conditions from \Eqref{eq:qc_boundary},
\begin{align}
q_{c,2}(t)&= \begin{cases}
		  0,             &\quad t \le 0 \\
		   f_c(t),       &\quad 0 < t < t_f/2 \\
		  -f_c(t_f-t)+d, &\quad t_f/2 < t < t_f \\
		  d,             &\quad t \ge t_f
		 \end{cases}.
\end{align}
To calculate $q_{0,2}(t)$, it is convenient to define $f_u(t)=1/\omega_0^2 f_c^{''}(t_f-t)$ where
\begin{align}	 
f_u(t)&= \frac{14 d}{3 \omega_0^2 t_f^2 }\left(2 \frac{t}{t_f}-1\right)^{1/3}.
\end{align}	
Similarly we first consider $f_u$ on $(t_f/2,t_f)$ and map appropriately to $(0,t_f/2)$, and obtain
\begin{align}
q_{0,2}(t)&= \begin{cases}
		  0, 					  &t \le 0 \\
		  f_c(t) + f_u(t_f-t), 	  &0 < t < t_f/2 \\
		 -f_c(t_f-t) -f_u(t) + d,  &t_f/2 < t < t_f \\
		  d,             		  &t \ge t_f
		 \end{cases},
\end{align}
that has discontinuities at $t=0,t=t_f/2$ and $t=t_f$.
These jump-points may be difficult to implement practically, so we smooth $q_{c,2}(t)$ in a time-interval of length $t_T$ around the jump points, using polynomial interpolation.
Later we will show that high performance can be achieved even with this smoothing process, and that the exact time-interval chosen is not critical to robust and high-fidelity transport.
Using \Eqref{eq:aux} we can now calculate $q_{0,2}(t)$,
\begin{align}
q_{0,2}(t)&= \begin{cases}
		  p_0''(t)/\omega_0^2  + p_0(t),  &0   \le t <   t_0 \\
		  f_c(t) + f_u(t_f-t), 		     &t_0 \le t \le t_1 \\
		  p_1''(t)/\omega_0^2  + p_1(t),  &t_1 <   t <   t_2 \\
		  -f_c(t_f-t) -f_u(t) + d,        &t_2 \le t \le t_3 \\
		  p_2''(t)/\omega_0^2  + p_2(t),  &t_3 \le t \le t_f
		 \end{cases},
\end{align}	
where $t_0=t_T$, $t_1=(t_f-t_T)/2$, $t_2=(t_f+t_T)/2$ and $t_3=t_f-t_T/2$.
By design, the polynomials $p_j(t)$ are matched to $f_c(t)$ and $-f_c(t)$ on the appropriate boundary points.
Throughout this paper we choose $t_T=t_f/8$.
An example of $q_{c,2}(t)$ and $q_{0,2}(t)$ can be seen in Fig. \ref{fig_2_qc_and_q0} (a) and (b), respectively (green dashed lines).

{\bf Quasi-optimal classical function $\bm{q_{c,3}(t)}$:}
Lastly we use a quasi-optimal classical auxiliary function as described in \cite{peterDemonstrationQuantumBrachistochrones2021}.
This auxiliary function is also derived via transport time-minimization in \cite{chenOptimalTrajectoriesEfficient2011}.
One motivation for this function is to consider the classical version of the particle transport problem.
The intuitive optimal strategy in this case is to maximally accelerate the particle during the first half of the transport, and maximally decelerate the particle in the second half of the transport.
In \cite{peterDemonstrationQuantumBrachistochrones2021} a sudden initial and final displacement of the potential is also included; we omit this since these displacements will occur within the smoothing intervals.
We call this auxiliary function the quasi-optimal classical function, and define it by
\begin{align}\label{eq:quasi-opt-qc}
q_{c,3}(t)&= \begin{cases}
		  2d \left(\frac{t}{t_f}\right)^2,          &0 \le t < \frac{t_f}{2} \\
		  d\left[1- 2\left(\frac{t}{t_f}-1\right)^2\right], &\frac{t_f}{2} < t \le t_f
		 \end{cases}.
\end{align}
We obtain $q_{0,3}$ using \Eqref{eq:aux}, giving
\begin{align}\label{eq:quasi-opt-q0}
q_{0,3}(t)&= \begin{cases}
		  0, &t \le 0 \\
		  2d \left[ \left(\frac{t}{t_f}\right)^2 + \frac{2}{\omega_0^2 t_f^2} \right],              &0 < t < \frac{t_f}{2} \\
		  -2d \left( \frac{t}{t_f} - 1 \right)^2 - d \left( \frac{4}{\omega_0^2 t_f^2} - 1 \right), &\frac{t_f}{2} < t < t_f \\
		  d ,&t \ge t_f 
		 \end{cases}.
\end{align}
We perform the same smoothing procedure as in the previous section.
In the next section we investigate the impact smoothing has on fidelity for this trajectory.
Examples of $q_{c,3}(t)$ and $q_{0,3}(t)$ can be seen in Fig. \ref{fig_2_qc_and_q0} (a) and (b) respectively (solid red lines).


\subsection{Derivation of eSTA control functions}\label{subsection:eSTA_Q}

To derive the eSTA control function, we must calculate $\vec{\epsilon}$ in \Eqref{eq:epsilon}.
In Fig. \ref{fig_1_schematic} (b) the improved eSTA control vector is shown schematically, with $\vec{\lambda}_s=\vec{\lambda}_0+\vec{\epsilon}$.
$\vec{\lambda}_0$ characterizes the STA control function, while $\vec{\epsilon}$ parameterizes the eSTA correction.
The purpose of this distinction is to highlight how eSTA improves the control of a system, starting with an idealized STA system with control vector $\vec{\lambda}_0$.

Now we wish to parameterize explicitly the eSTA modification and, without loss of generality, we can simplify our chosen parameterization by choosing $\vec{\lambda}_0 = \vec{0}$.
This allows us to define the new improved eSTA control function $Q_{j}(\vec{\epsilon},t)$ as the sum of the original STA control function and a second function $\Delta q_{0,j}(\vec{\epsilon},t)$,
\begin{align}\label{eq:Q}
Q_{j}(\vec{\epsilon},t) = q_{0,j}(t) + \Delta q_{0,j}(\vec{\epsilon},t).
\end{align}
We now have the freedom to define $\Delta q_{0,j}$ and the $\vec{\epsilon}$ parameterization in any manner that is convenient, provided that $Q_{j}(\vec{\epsilon},t)$ remains consistent with the boundary conditions of $q_{0,j}(t)$.

We define $\vec{\epsilon}$ by values that $ \Delta q_{0,j}$ takes for equally spaced points in time during the transport.
It is then convenient to set $ \Delta q_{0,j}$ to be a polynomial
\begin{align}
\Delta q_{0,j}(\vec{\epsilon},t)=\sum_{l=0}^{L+5} b_l t^l,
\end{align}
that satisfies
\begin{align}\label{eq:Delta_q0_conds}
\Delta q_{0,j}(t') &= 0, \quad t=0,t_f \nonumber \\
\frac{\partial^n}{\partial t^n }\Delta q_{0,j}(t') &= 0, \quad t=0,t_f \text{ and } n=1,2, \nonumber \\
\Delta q_{0,j}\left(\frac{l\: t_f}{L}\right) &= \epsilon_l, \quad l=1,\dots,L.
\end{align}
We choose $L=8$ in this paper as a good compromise between numerical implementation and optimization freedom, further details can be found in \cite{whittyQuantumControlEnhanced2020}.

To implement eSTA for a given trajectory, we calculate $\vec{\epsilon}$ using \Eqref{eq:epsilon}.
The states $\ket{\chi_n (s)}$ are known analytically from \Eqref{eq:transport_modes}, and so the integrals $G_n$ and $K_n$ can be calculated for each $n$.
We choose $N=4$ for all the results presented in this paper as terms beyond $N=4$ do not have an impact on the resulting fidelities or robustness, for this physical setting.

\begin{figure*}[ht]
\begin{center}
(a) \includegraphics[width=\imageScale\columnwidth]{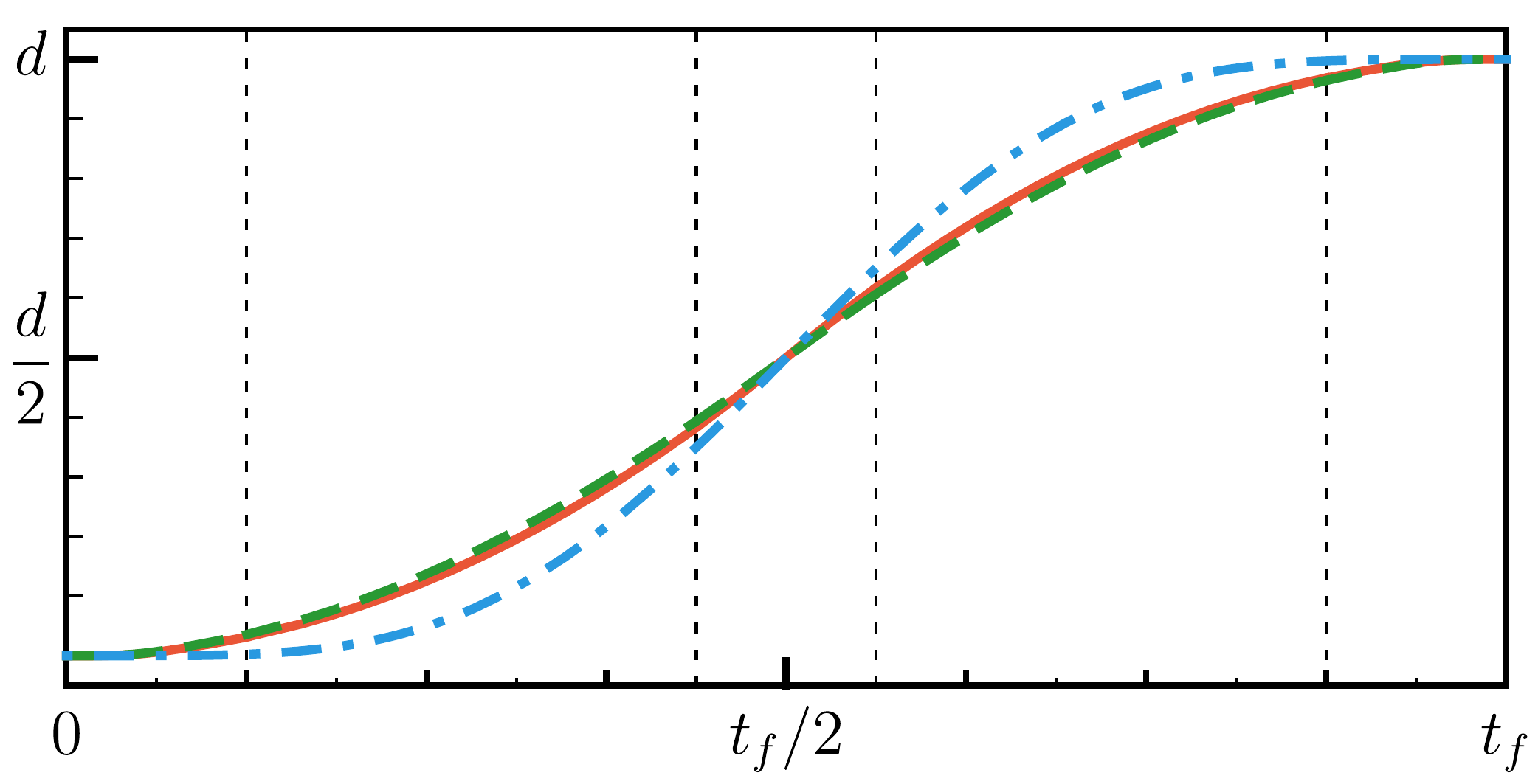}
(b) \includegraphics[width=\imageScale\columnwidth]{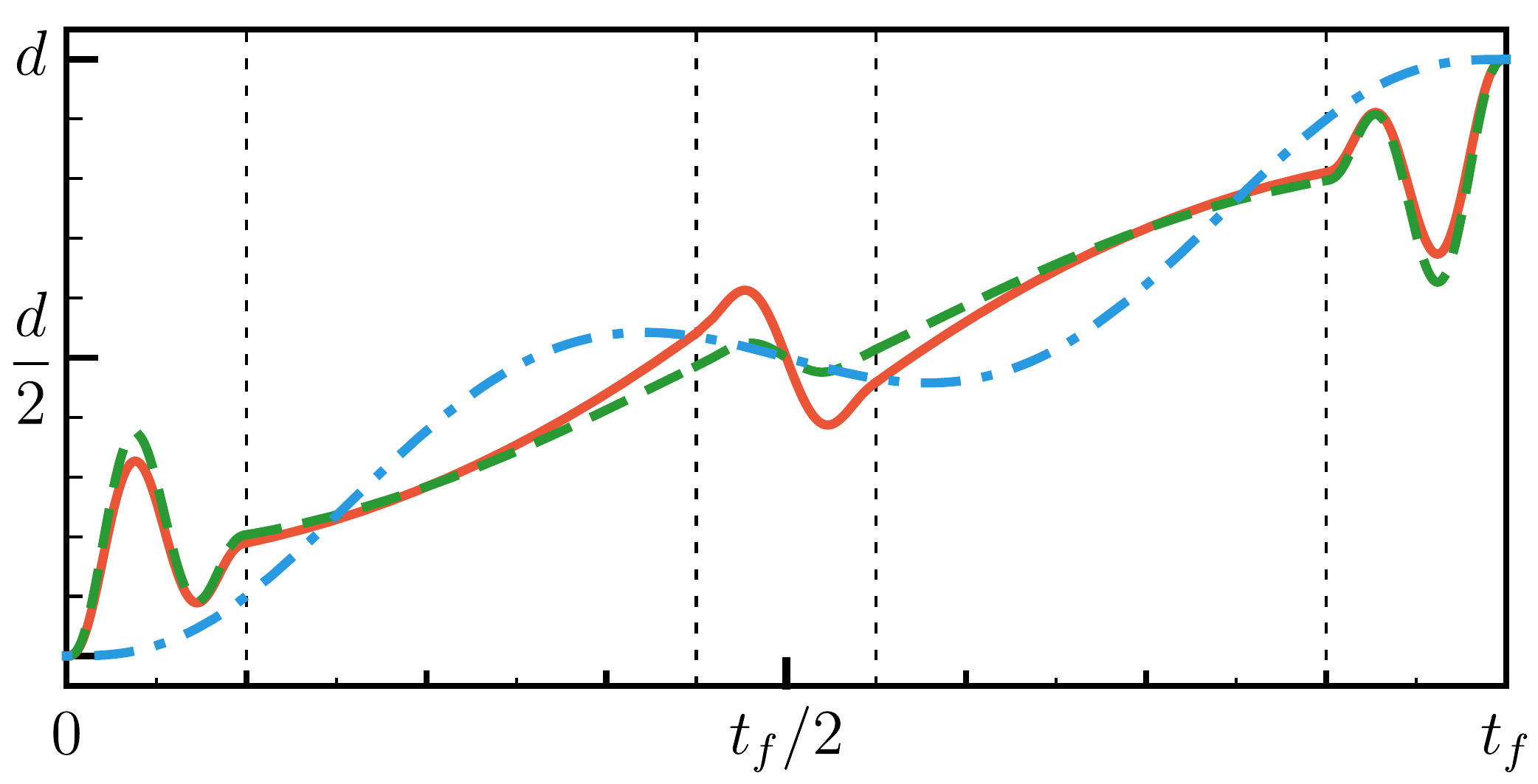}\\
(c) \includegraphics[width=\imageScale\columnwidth]{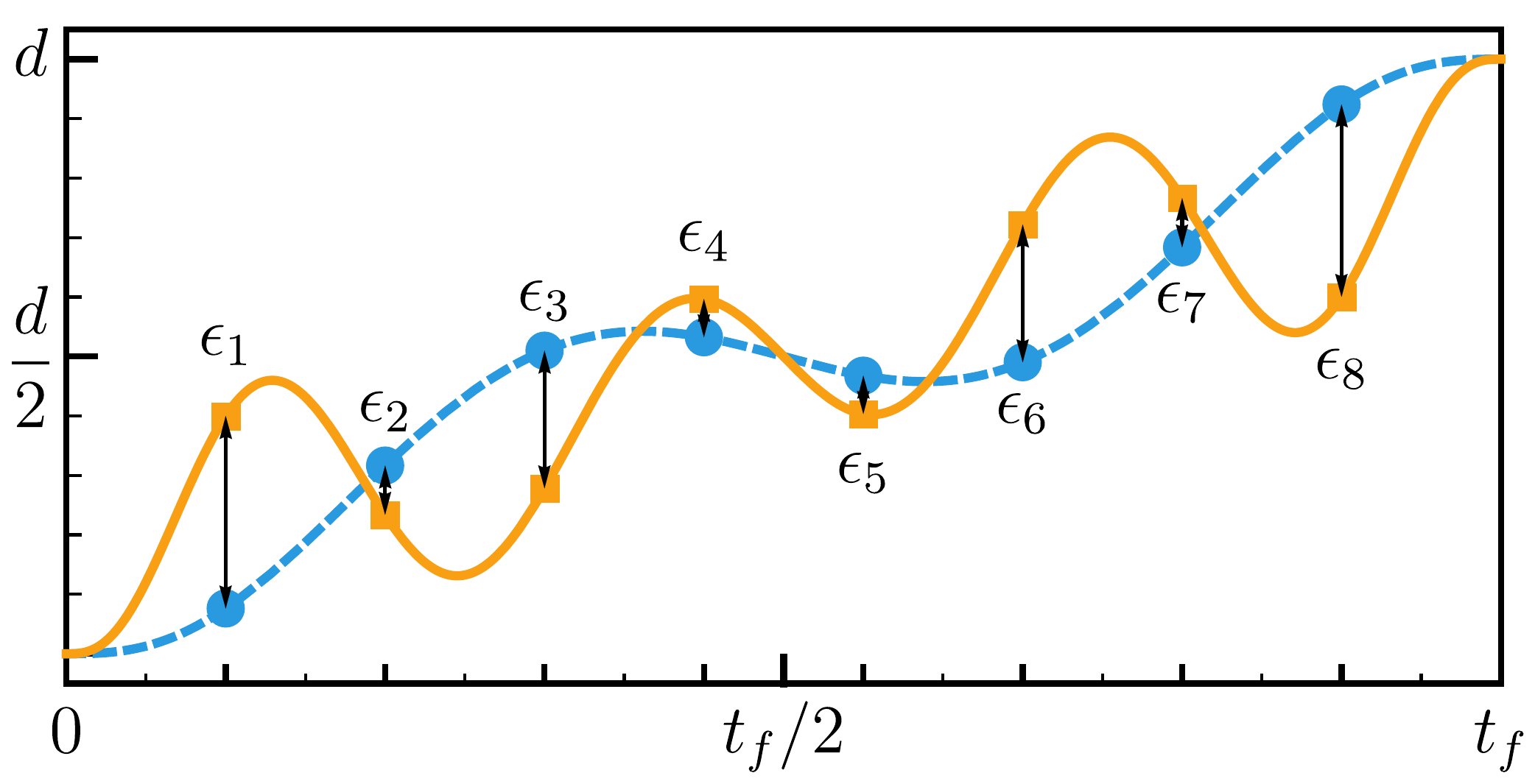}
(d) \includegraphics[width=\imageScale\columnwidth]{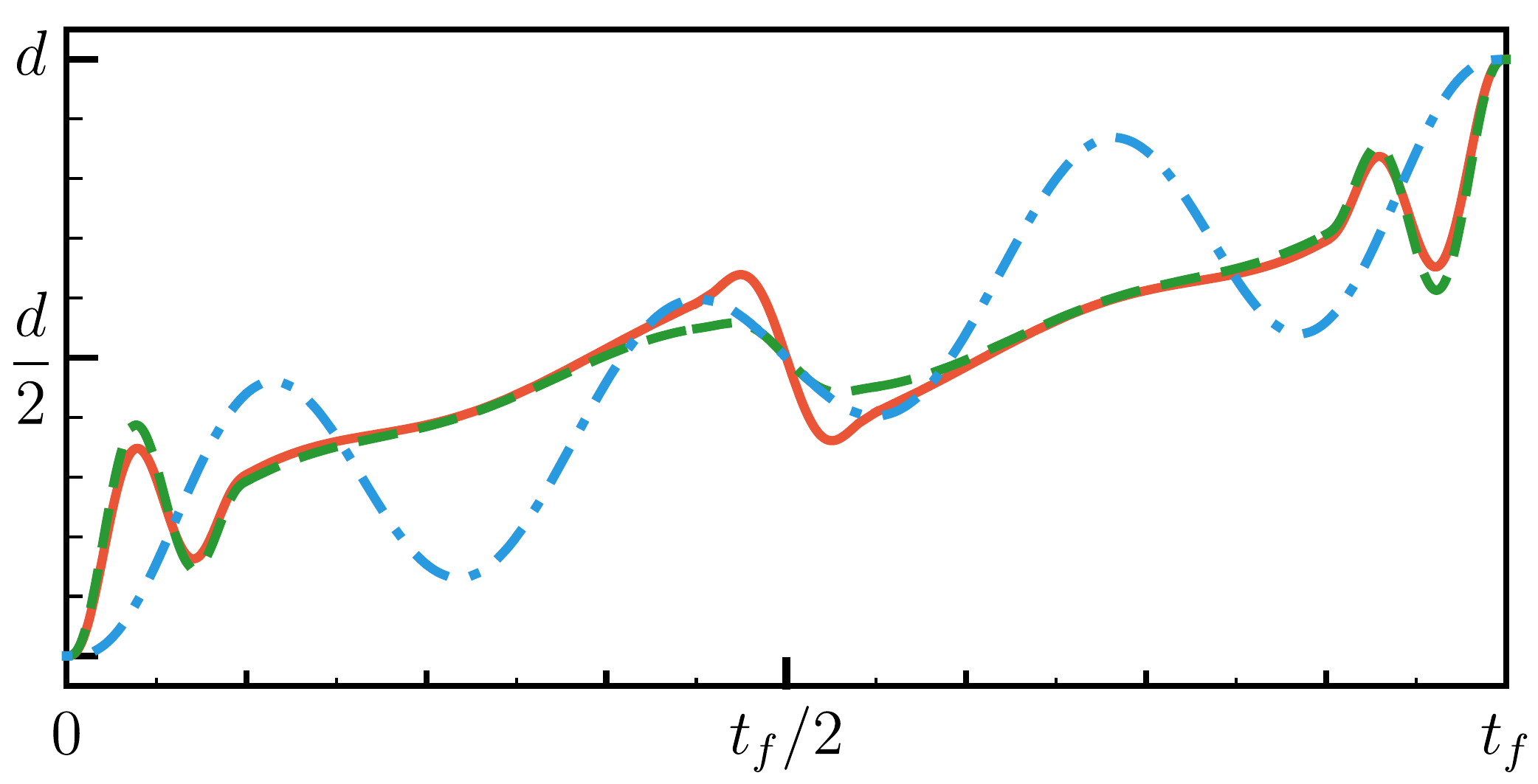}
\end{center}
\caption{\label{fig_2_qc_and_q0} 
Plot of STA and eSTA control functions.
(a) Plot of the auxiliary functions, $q_{c,1}(t)$ (dot-dashed blue), $q_{c,2}(t)$ (dashed green) and $q_{c,3}(t)$ (solid red). In each of these plots $t_f/\tau=0.8$ and $d=\lambda/2\sigma$ (one lattice site).
(b) Corresponding plots of the STA transport functions, $q_{0,1}(t)$ (dot-dashed blue), $q_{0,2}(t)$ (dashed green) and $q_{0,3}(t)$ (solid red).
(c) Example of the vector components of the eSTA correction $\vec{\epsilon}$, with $q_{0,1}$ (dashed blue) and $Q_{1}$ (solid orange) shown for $t_f/\tau=0.8$ and $d=\lambda/2\sigma$.
(d) Examples of the eSTA control functions.
The vertical dashed lines in plots (a) and (b) indicate the smoothing boxes of length $t_f/8$ about the discontinuities in $q_{c,2}$, $q_{0,2}$, $q_{c,3}$ and $q_{0,3}$.
}
\end{figure*}

Both $\Delta q_{0,j}$ and $\vec{\epsilon}$ are illustrated in Fig. \ref{fig_2_qc_and_q0} (c), with the STA trajectory $q_{0,1}(t)$ (dot-dashed blue line) and the improved eSTA trajectory $Q_{1}(\vec{\epsilon},t)$ (solid orange line) shown for $t_f/\tau=0.8$.
The magnitude of the $\vec{\epsilon}$ components $\epsilon_l$ are shown explicitly as changes to the original STA trajectory at the times $l \:t_f/L$, where $l=1,\dots,L$ and $L=8$.

\subsection{Fidelities for STA and eSTA schemes}

We investigate the fidelities using STA and eSTA by numerically simulating the Schr\"odinger equation with the lattice Hamiltonian from \Eqref{eq:H_s}, for short transport times.
We choose $m=133$ amu ($^{133}$Cs),  $\lambda=866$ nm and $\alpha=150$, motivated by the physical values stated for lattice transport near the quantum speed limit in \cite{peterDemonstrationQuantumBrachistochrones2021}.
This choice of units correspond to a natural time unit of $\tau=20\, \mu s$, where $\tau$ is approximately the quantum speed limit for this transport given in \cite{peterDemonstrationQuantumBrachistochrones2021}.

\begin{figure}[t!]
\begin{center}
\includegraphics[width=\imageScale\columnwidth]{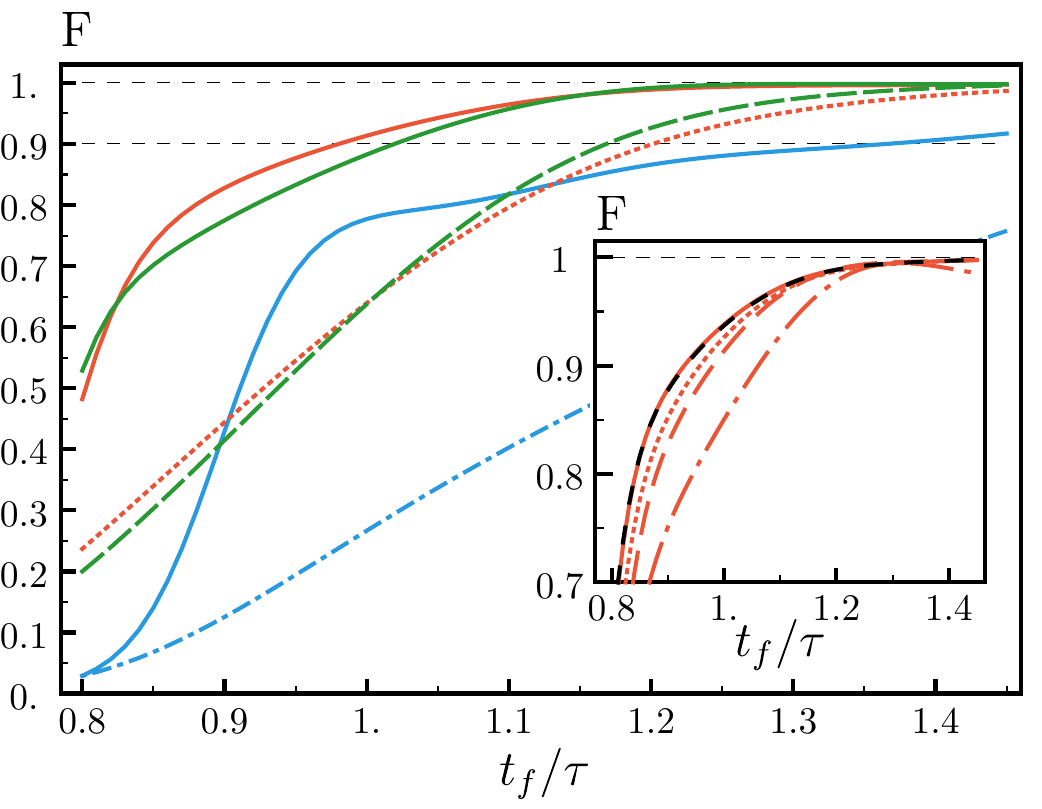}
\end{center}
\caption{\label{fig_3_fid} Fidelity $F$ versus final time $t_f$, for $\delta=0$.
The fidelities for the STA trajectories are given by the broken lines; $q_{0,1}(t)$ (dot-dashed blue), $q_{0,2}(t)$ (dashed green) and $q_{0,3}(t)$ (dotted red).
The corresponding eSTA optimized fidelities are solid lines; $Q_{1}(t)$ (blue), $Q_{2}(t)$ (green) and $Q_{3}(t)$ (red).
Inset of (a): Fidelity $F$ versus final time $t_f$ using $Q_3$ with different smoothing options.
The solid-red line uses a fully discontinuous $q_{c,3}$, and the dashed-black line on top of it uses $q_{c,3}$ with only the center discontinuity smoothed over a $t_T=t_f/16$ interval.
The dotted-red line uses $q_{c,3}$ smoothed in an interval of length $t_T=t_f/16$ around the three discontinuities and the dashed-red line uses $q_{c,3}$ smoothed in an interval of length $t_T=t_f/8$ around the three discontinuities.
}
\end{figure}

We first consider the fidelity of the STA trajectories $q_{0,1}(t)$, $q_{0,2}(t)$ and $q_{0,3}(t)$, and the results are shown in Fig. \ref{fig_3_fid} (a).
The STA trajectories based on quasi-optimal solutions [$q_{0,2}(t)$ and $q_{0,3}(t)$] perform better than a simple polynomial ansatz [$q_{0,1}(t)$].
We now fix a reference fidelity $F_R=0.9$, and we see that both $q_{0,2}(t)$ and $q_{0,3}(t)$ have $F>F_R$ for $t_f/\tau \gtrapprox 1.2$, while $t_f/\tau \gg 1.5$ is required for $q_{0,1}(t)$.
Since the performance of STA is already optimal for $t_f/\tau \approx 1.45$, we will focus on the region $t_f/\tau < 1.5$.

The fidelities for the three eSTA optimized trajectories $Q_{1}(t)$, $Q_{2}(t)$ and $Q_{3}(t)$ are also shown in Fig. \ref{fig_3_fid}(a).
They show improvement over their corresponding STA trajectories for the transport times considered.
As with the STA trajectories, the eSTA trajectories also show that the quasi-optimal solutions out-perform the polynomial ansatz.
Furthermore, the eSTA trajectories produce higher fidelities for shorter times than the STA trajectories; $Q_{0,2}(t)$ has $F>F_R$ for $t_f/\tau \gtrapprox 1.025$ and $Q_{0,3}(t)$ has $F>F_R$ for $t_f/\tau \gtrapprox 0.98$. 

As a side remark, we investigated whether the smoothing we performed on the trajectories had a significant impact on fidelity.
As an example, in inset of Fig. \ref{fig_3_fid} the fidelities for different smoothing possibilities are shown for $Q_3(t)$.
It was found that the smoothing interval $t_T$ was not critical in obtaining high fidelities.
While the highest fidelity is obtained for the fully discontinuous trajectory (dashed line), very similar results are found using a smoothing interval even as large as $t_T=t_f/8$ (dashed line).
An alternative approach is to use a fully discontinuous $q_{c,3}(t)$ and perform a smoothing procedure on the resulting eSTA trajectory $Q_{0,3}(t)$. However, this was found to give poorer performance, as shown in the inset of Fig. \ref{fig_3_fid} (dot-dashed line).
Thus, for the results in this paper the STA trajectories $q_{c,j}(t)$ were smoothed using $t_T=t_f/8$, and then the eSTA trajectories $Q_j(t)$ were calculated.

\section{eSTA Control Function Deviation \label{sect_C_Q}}

Before starting with an examination of the robustness of eSTA, we will first examine the related question of how much the eSTA control function deviates when parameters within the potential are slightly changed. 
We define the deviation $\cC_Q$ of the control function $Q$.
The motivation for examining $\cC_Q$ is an expectation that if the control function does not depend strongly on a specific parameter of the potential, then this could result in stability concerning systematic errors in that parameter.

In detail, we define the deviation $\cC_Q$ as
\begin{align}\label{eq:C_Q}
\cC_Q := \lim_{\delta \rightarrow 0}\frac{1}{\delta} \fnorm{  Q(\delta)-Q(0) }
=
\fnorm{ \frac{\partial}{\partial \delta}Q\Bigr|_{\delta=0}}.
\end{align}
where $\delta$ is some variation of the potential.

There is freedom in the choice of norm in \Eqref{eq:C_Q}, in the following we will use the $L_1$ norm
\begin{align}
\fnorm{Q(\delta)} = \int_0^{t_f} ds \fabs{Q(\delta,s)}.
\end{align}

As in \Eqref{eq:Q}, we assume that the solution to the eSTA system with Hamiltonian $\cH(\delta)$ takes the form
\begin{align}
Q(\delta, t) = q_0(\delta,t) + \Delta q_0 \left[ \vec{\epsilon}(\delta),t\right],
\end{align}
where $q_0(\delta,t)$ is the STA control function that solves the approximate STA system, and $\Delta q_0$ is a polynomial as defined in \Eqref{eq:Delta_q0_conds}. 
We set $ \Delta q_0 \left[ \vec{0},t\right] = 0$ and assume $\Delta q_0$ does not depend on the STA control function $q_0(\delta,t)$.
We also have that
\begin{align}
\frac{\partial Q}{\partial \delta} \Bigr|_{\delta=0}
&=
\frac{\partial q_0}{\partial \delta} \Bigr|_{\delta=0}
+
\frac{\partial }{\partial \delta} \Delta q_0 (\vec{\epsilon}(0),t) \Bigr|_{\delta=0}
\nonumber \\
&=
\frac{\partial q_0}{\partial \delta} \Bigr|_{\delta=0}
+
\sum_{j=1}^{N}
\frac{\partial \Delta q_0}{\partial \epsilon_j}(\vec{\epsilon}(0),t) \: \frac{\partial \epsilon_j}{\partial \delta}\Bigr|_{\delta=0}.
\end{align}
Using the definition of $\cC_Q$ in \Eqref{eq:C_Q} we obtain
\begin{align}\label{eq:C_Q_explicit}
\cC_Q :&= 
\fnorm{
\frac{\partial q_0}{\partial \delta} \Bigr|_{\delta=0}
+
\sum_{j=1}^{N}
\frac{\partial \Delta q_0}{\partial \epsilon_j}(\vec{\epsilon}(0),t)\:
\frac{\partial \epsilon_j}{\partial \delta}\Bigr|_{\delta=0}
}
\end{align}
By using the eSTA formalism, we can calculate $\frac{\partial \epsilon_j}{\partial \delta}$ explicitly as shown in Appendix \ref{app:s_Q}.
Starting from \Eqref{eq:C_Q_explicit}, we can also derive an upper bound of the quantity $C_Q$:
\begin{align}\label{eq:C_Q_upper_bound}
\cC_Q &\le
\fnorm{
\frac{\partial q_0}{\partial \delta} \Bigr|_{\delta=0}
}
+
\sum_{j=1}^{N}
\fnorm{
\frac{\partial \Delta q_0}{\partial \epsilon_j}(\vec\epsilon(0),t)
}
\fabs{
\frac{\partial \epsilon_j}{\partial \delta}
}_{\delta=0}
\nonumber \\
&\le
\fnorm{
\frac{\partial q_0}{\partial \delta} \Bigr|_{\delta=0}
}
+
\left[
\max_j 
\fnorm{
\frac{\partial \Delta q_0}{\partial \epsilon_j}(\vec\epsilon(0),t)
}
\right]
\sum_{j=1}^{N}
\fabs{
\frac{\partial \epsilon_j}{\partial \delta}
}_{\delta=0}.
\end{align}
The first term $\partial q_0 / \partial \delta$ is the deviation of the STA trajectory.
The second term is a measure of the eSTA dependence with respect to the control function $Q$, and this is the term we wish to investigate.
Note that the calculation of $\cC_Q$ can be done fully analytically as shown in Appendix \ref{app:s_Q}, and so requires far less computation than the numerical derivative of the fidelity which we will consider in the next section.
We highlight here that $\cC_Q$ offers potential as a tool to evaluate and classify possible eSTA trajectories in lieu of full numerical treatment.
As an example, we consider a correlated error in the lattice amplitude $U_0$ and wavenumber $k_0$:
\begin{align}\label{eq:a_err}
V_{\text{err}}^c(x,t) = U_0(1+\delta) \operatorname{sin}^2 \left\{k_0 \frac{\left[x-Q_{j}(t)\right]}{\sqrt{(1+\delta)}} \right\},
\end{align}
such that $\omega =\omega_0= \sqrt{\frac{2 U_0}{m}} k_0$ is kept constant, hence the STA trajectories do not depend on $\delta$.
Systematic errors in the lattice amplitude or wavenumber alone will be considered in following sections.
This error potential allows us to focus on applying $\cC_Q$ to eSTA control functions, since the STA trajectories $q_{0,j}$ do not depend on $\delta$.

\begin{figure}[t]
\begin{center}
\includegraphics[width=\imageScale\columnwidth]{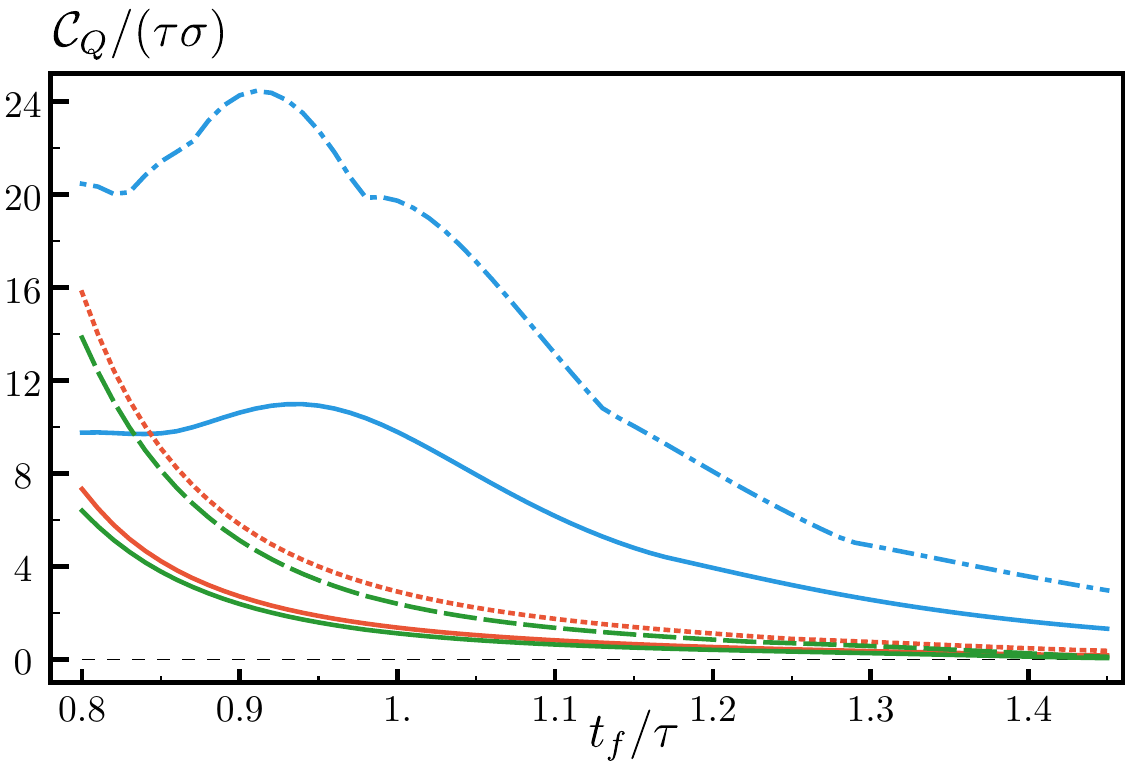}
\end{center}
\caption{\label{fig_analytic_sens_qcClass_qcMinAnharm} 
Control function deviation $\cC_Q$ versus $t_f$ for correlated systematic error.
Trajectories $Q_{1}(t)$ (solid blue), $Q_{2}(t)$ (solid green) and $Q_{3}(t)$ (solid red).
Upper bound of $\cC_Q$: trajectories $Q_{1}(t)$ (dot-dashed blue), $Q_{2}(t)$ (dashed green) and $Q_{3}(t)$ (dotted red).
}
\end{figure}

The corresponding results of $\cC_Q$ can be seen in Fig. \ref{fig_analytic_sens_qcClass_qcMinAnharm}.
We find that the trajectories $Q_2$ and $Q_3$ (solid lines) show a lower deviation with changes in $\delta$ than the trajectory $Q_1$.
The upper bound on $\cC_Q$ from \Eqref{eq:C_Q_upper_bound} is also shown and we see that the upper bound can be also used for classifying the different schemes.
From these results, one would expect that the trajectories $Q_2$ and $Q_3$ are more stable concerning a systematic error $\delta$.
This will be examined in the next section in detail and will be shown to be the case. 

\section{Robustness to Systematic Errors \label{sect_robustness_systematic}}\label{section:sys_err_sens}

In this section we compare the robustness of the eSTA and STA trajectories by considering how the fidelity changes under three systematic errors in the lattice potential.

We first consider the correlated error $V_{\text{err}}^c$ introduced in the last section for a specific final time and then define a sensitivity $S$ for a given trajectory.
We compare eSTA to STA using three systematic errors: the correlated error $V_{\text{err}}^c$ and two further errors, an error in the lattice amplitude $V_{\text{err}}^A$ and an error in the lattice wavenumber $V_{\text{err}}^k$.
The amplitude and wavenumber errors can occur in the physical implementation of lattice potentials \cite{luFastShuttlingParticle2018}.

We define $\delta$ to be the strength of the systematic error, and define two further error potentials:
The lattice amplitude error potential is given by
\begin{align}\label{eq:amp_err}
V_{\text{err}}^A(x,t) = U_0(1+\delta) \operatorname{sin}^2 \left\{ k_0 \left[ x-Q_{j}(t) \right] \right\},
\end{align}
and the wavenumber error potential is
\begin{align}\label{eq:omega_err}
V_{\text{err}}^k(x,t) = U_0 \operatorname{sin}^2 \left\{ k_0 \sqrt{1+\delta} \left[ x-Q_{j}(t) \right] \right\},
\end{align}
with $\omega=\omega_0 \sqrt{(1+\delta)}$ in both cases.

\begin{figure}[t]
\raggedright
\includegraphics[width=\imageScale\columnwidth]{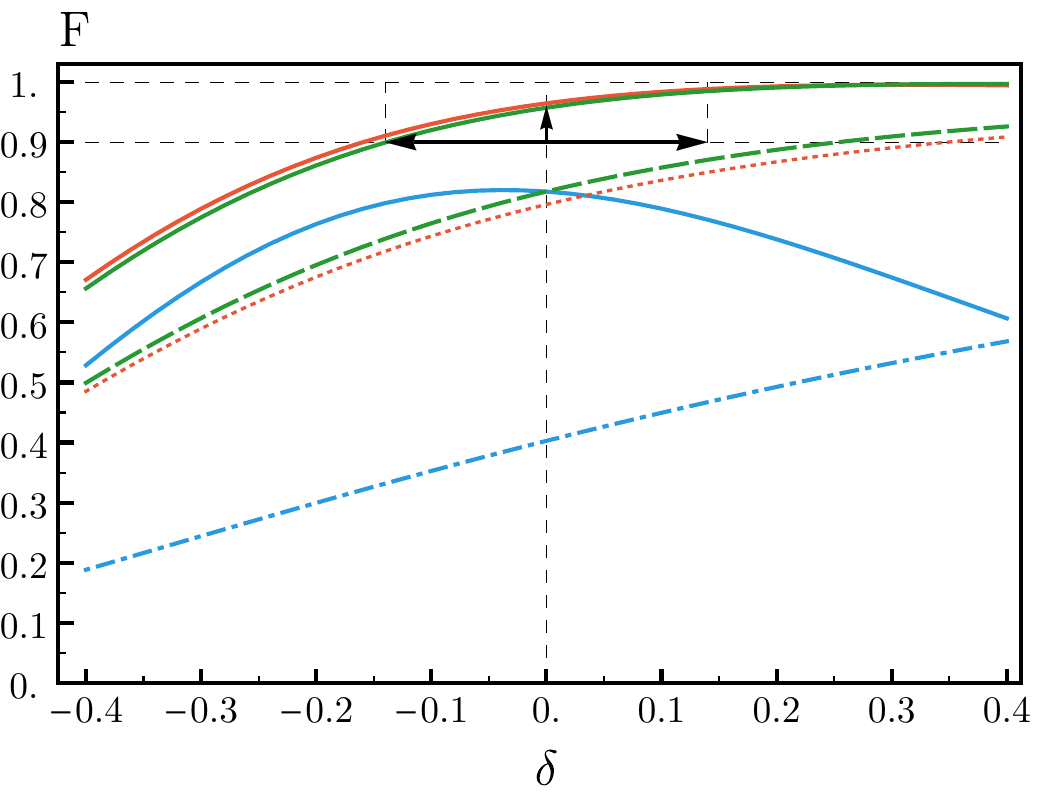} 
\caption{\label{fig_5} Fidelity at $t_f/\tau=1.1$ versus $\delta$.
The fidelities for the STA trajectories are shown, for $q_{0,1}(t)$ (dot-dashed blue), $q_{0,2}(t)$ (dashed green) and $q_{0,3}(t)$ (dotted red).
The corresponding eSTA optimized fidelities are $Q_{1}(t)$ (solid blue), $Q_{2}(t)$ (solid green) and $Q_{3}(t)$ (solid red).
The horizontal solid-black arrows show $\fabs{\delta}$, such that $F(\delta) > F_R=0.9$ for $Q_2$ (solid green).
The vertical solid-black arrow at $\delta=0$ is $F(\delta=0)-F_R$.
These quantities are used in the definition of the systematic error bound $\cB$, in \Eqref{eq:B}.
}
\end{figure}

As a first step, we consider the correlated error $V_{\text{err}}^c$ for a fixed final time of $t_f/\tau = 1.1$.

In Fig. \ref{fig_5}, the fidelity $F$ versus the error strength $\delta$ is shown for the three eSTA and three STA trajectories.
The STA trajectories $q_{0,2}$ and $q_{0,3}$ show similar fidelities about $\delta=0$, and the same is true for the eSTA trajectories $Q_2$ and $Q_3$.
We see significant higher fidelities of the eSTA schemes $Q_2$ and $Q_3$ over the STA schemes $q_{0,2}$ and $q_{0,3}$, respectively.
While the eSTA polynomial ansatz $Q_1$ has much higher fidelity than the STA $q_{0,1}$, $Q_1$ has fidelity $F<0.9$ for all $\delta$ in the range considered.

We see that there is no significant change in the fidelity for the eSTA trajectories in a neighborhood about $\delta=0$, and that even for larger values of $\delta$ the eSTA trajectories maintain their higher fidelity over their related STA trajectories.

We will show later that even $\partial F/\partial \delta$ at $\delta=0$ is smaller for the eSTA trajectories $Q_2$ and $Q_3$, than the STA trajectories 
$q_{0,2}$ and $q_{0,3}$ (see Fig. \ref{fig_6}).
This can also already be seen in Fig. \ref{fig_5}, as the slopes of the eSTA lines (solid green and red line) are less than those for the STA lines (dashed-green line and dotted-red line).

Note that for $V_{\text{err}}^c$, increasing $\delta$ corresponds with deepening of the lattice.
As the lattice is deepened we find increased fidelity and stability for both $Q_2$ and $Q_3$, and $q_{0,2}$ and $q_{0,3}$.

\subsection{Systematic Error Sensitivity} 

To examine the robustness of eSTA quantitatively for the three types of systematic errors we define the sensitivity
\begin{align}\label{eq:S_systematic}
S := \fabs{\frac{\partial F}{\partial \delta}\Bigr|_{\delta=0}}.
\end{align}
A smaller value of $S$ corresponds to a more robust protocol, i.e. less sensitivity to the error induced by $\delta$.
We evaluate $S$ for the different errors and trajectories numerically around $\delta=0$ by simulating the full transport.
Note that using time-dependent perturbation theory, $S$ can be expressed as
\begin{align}\label{eq:S_systematic_calc}
S = \frac{2}{\hbar}\Big| 
\text{Im} \Big\{
\int_0^{t_f} dt \,
&\braXket{\Psi_T(t)}{\frac{\partial H}{\partial \delta}\Bigr|_{\delta=0} }{\Psi_0(t)}
\nonumber \\
\times &\braket{\Psi_0(t_f)}{\Psi_T(t_f)}
\Big\}
\Big|,
\end{align}
where $\ket{\Psi_0 (0)}$ is the initial state, $\ket{\Psi_T (t_f)} $ is the target state and $\ket{\Psi_0 (t)}$ is the time-evolved solution of the Schr\"odinger equation.

In the following, we will evaluate both eSTA and STA for the three systematic errors stated previously.
Since we are interested in trajectories that give the highest fidelity, we restrict our focus to $Q_2$ and $Q_3$ ($q_{0,2}$, $q_{0,3}$ respectively).

\begin{figure}[t]
\begin{center}
(a) \includegraphics[width=\imageScale\columnwidth]{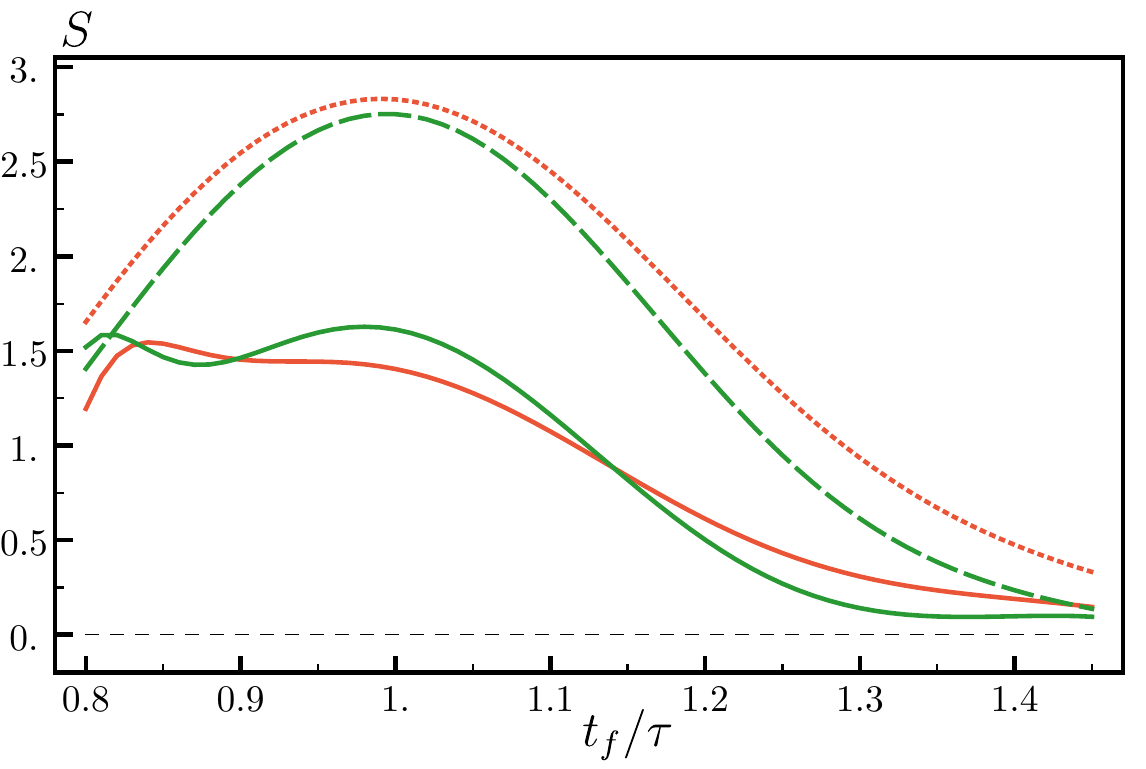}\\
(b) \includegraphics[width=\imageScale\columnwidth]{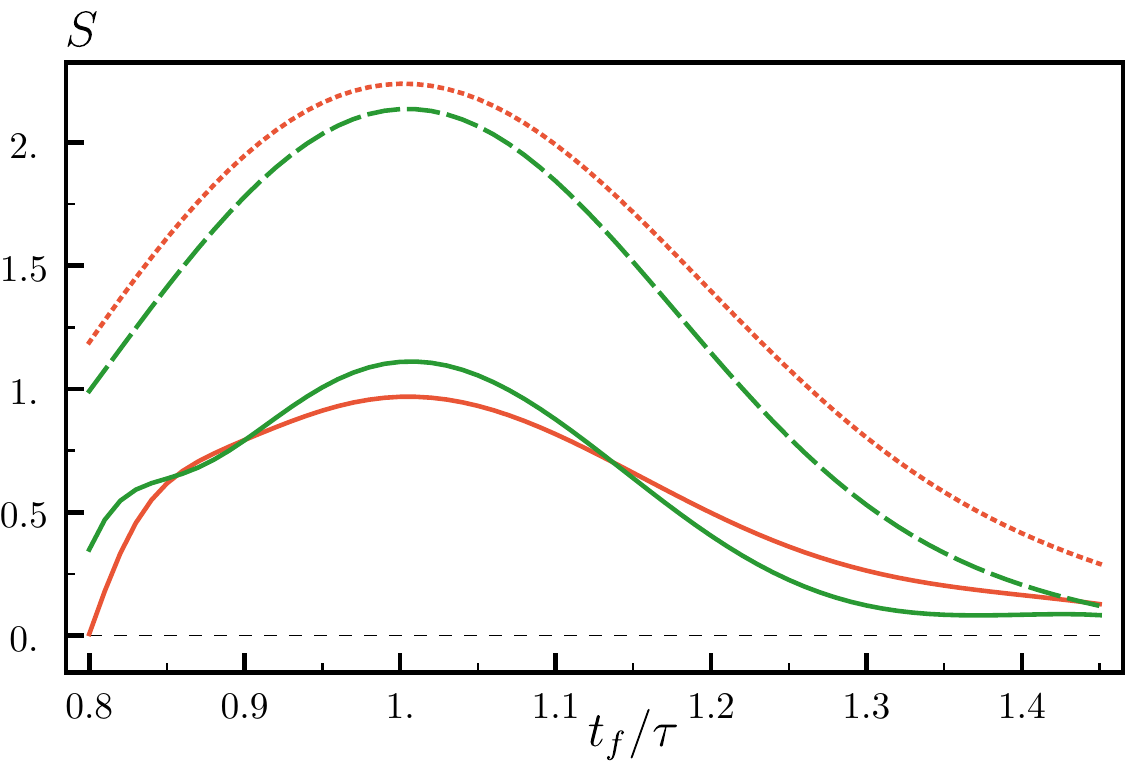}\\
(c) \includegraphics[width=\imageScale\columnwidth]{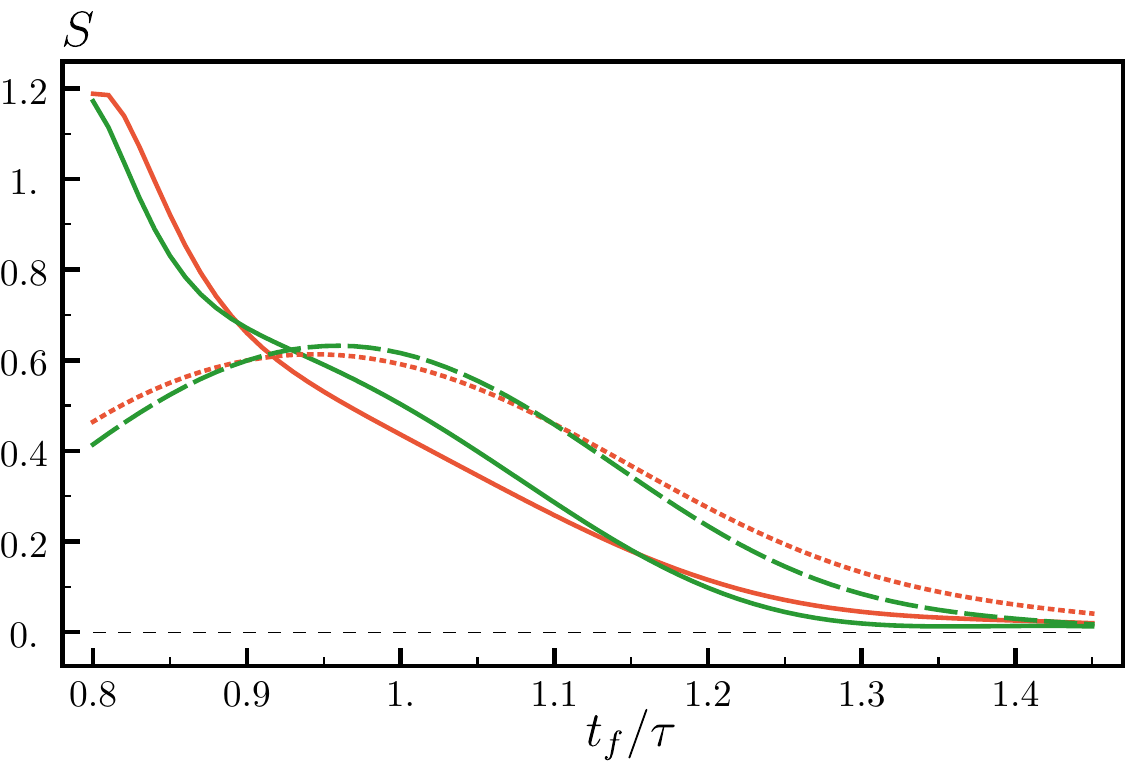}
\end{center}
\caption{\label{fig_6} 
Sensitivity $S$ for systematic errors versus $t_f$ for (a) amplitude error $V_{\text{err}}^A$, (b) wavenumber error $V_{\text{err}}^k$, (c) correlated error $V_{\text{err}}^c$.
eSTA trajectories: $Q_2$ (solid green) and $Q_3$ (solid red). STA trajectories: $q_{0,2}$ (dashed green) and $q_{0,3}$ (dotted red).
}
\end{figure}

In Fig. \ref{fig_6} (c) we consider the correlated error $V_{\text{err}}^c$. For $t_f/\tau \ge 0.95$, eSTA shows reduced sensitivity over STA.
Note that the fidelities are also higher in this $t_f$ range (see Fig. \ref{fig_3_fid}).
We also find that $Q_2$ and $Q_3$ in Fig. \ref{fig_6} (c) both agree qualitatively with their analytic $\cC_Q$ behavior in Fig. \ref{fig_analytic_sens_qcClass_qcMinAnharm}.

We show the sensitivity of eSTA and STA versus $t_f/\tau$ for $V_{\text{err}}^A$ in Fig. \ref{fig_6} (a) and $V_{\text{err}}^k$ in Fig. \ref{fig_6} (b).
For these errors, the eSTA trajectories (solid lines) generally have lower sensitivities than the STA trajectories (dashed and dotted lines).

If we first consider longer transport times $t_f$, $S$ is approaching zero for every trajectory.
This behavior is expected given the Adiabatic theorem; as $t_f$ approaches the adiabatic limit, small perturbations in the potential will have less impact on the instantaneous eigenstate of the system.
Thus, $F\to 1$, and $S\to 0$.

For very short final times ($t_f/\tau<1$), $S$ becomes a less useful description of robustness since the fidelity is rapidly decreasing for all trajectories.
Hence a more useful quantity would consider the fidelity $F$ and sensitivity $S$ together, and this motivates us to define a new quantity in the next section that we call the systematic error bound.

\subsection{Systematic Error bound} 

For practical implementation, we are interested in protocols that exceed a chosen threshold fidelity $F_R$ while also having stability against a systematic error within a certain bound.
To address these concerns, we now define a new quantity $\cB$ called the systematic error bound.

\begin{figure}[t]
\begin{center}
(a) \includegraphics[width=\imageScale\columnwidth]{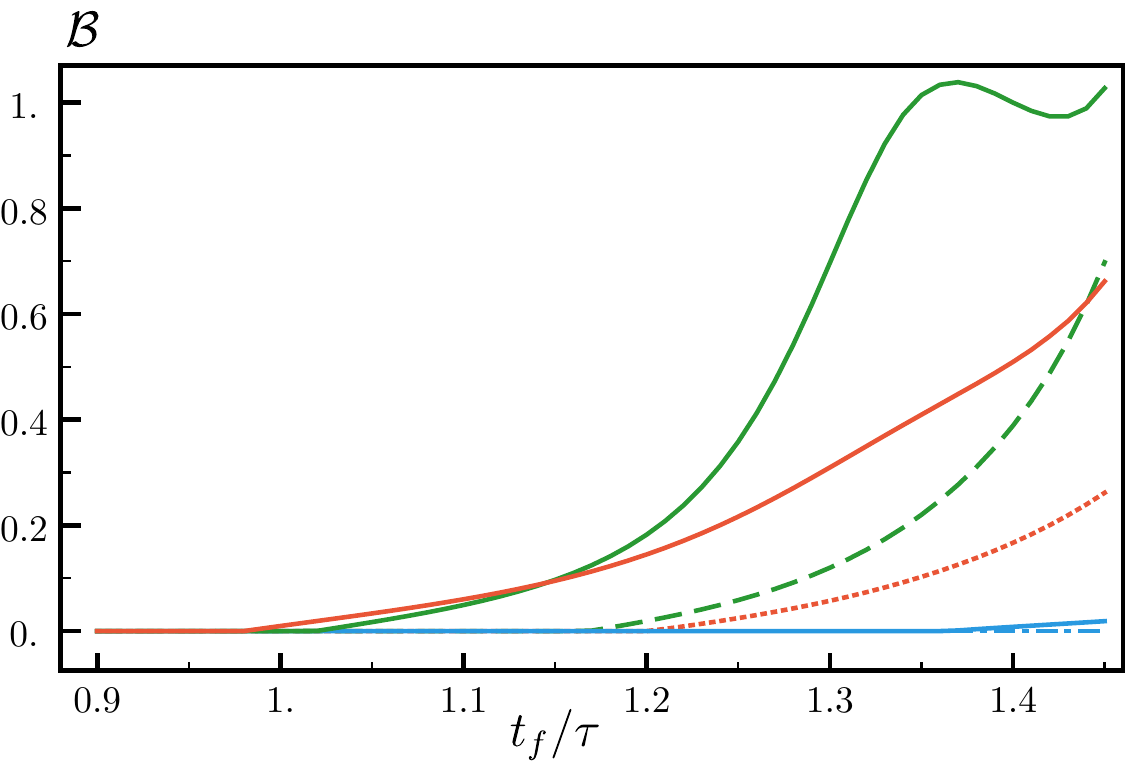}\\
(b) \includegraphics[width=\imageScale\columnwidth]{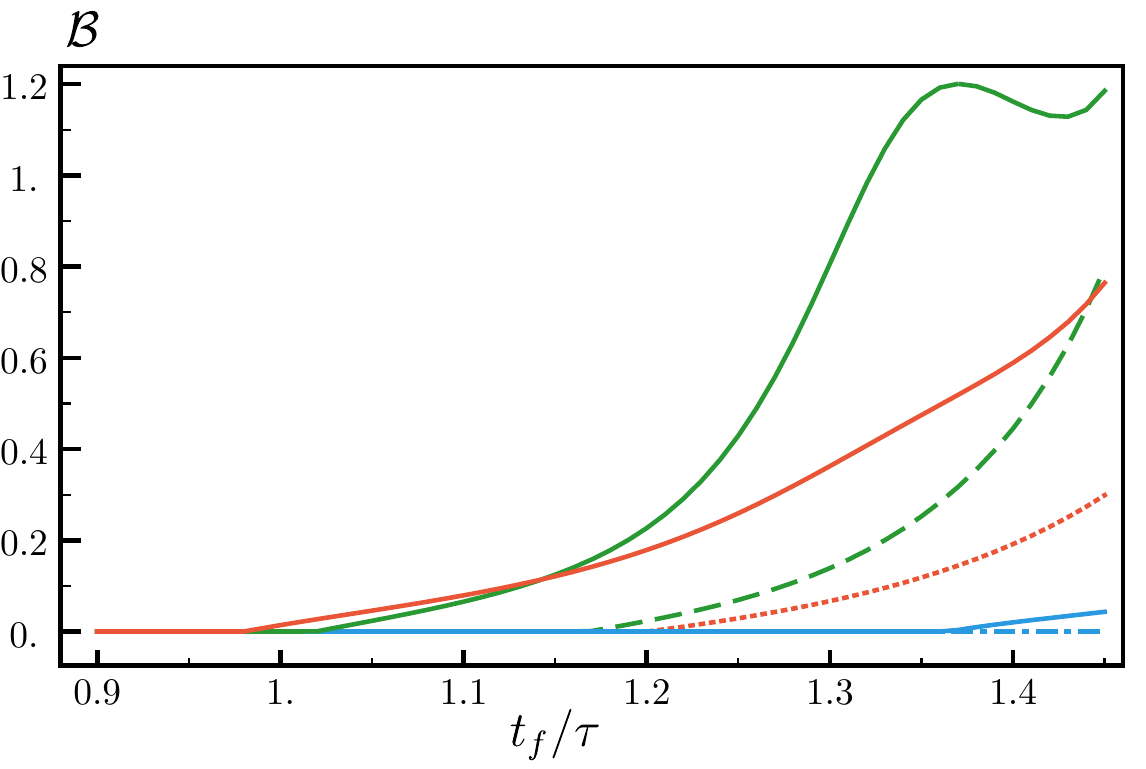}\\
(c) \includegraphics[width=\imageScale\columnwidth]{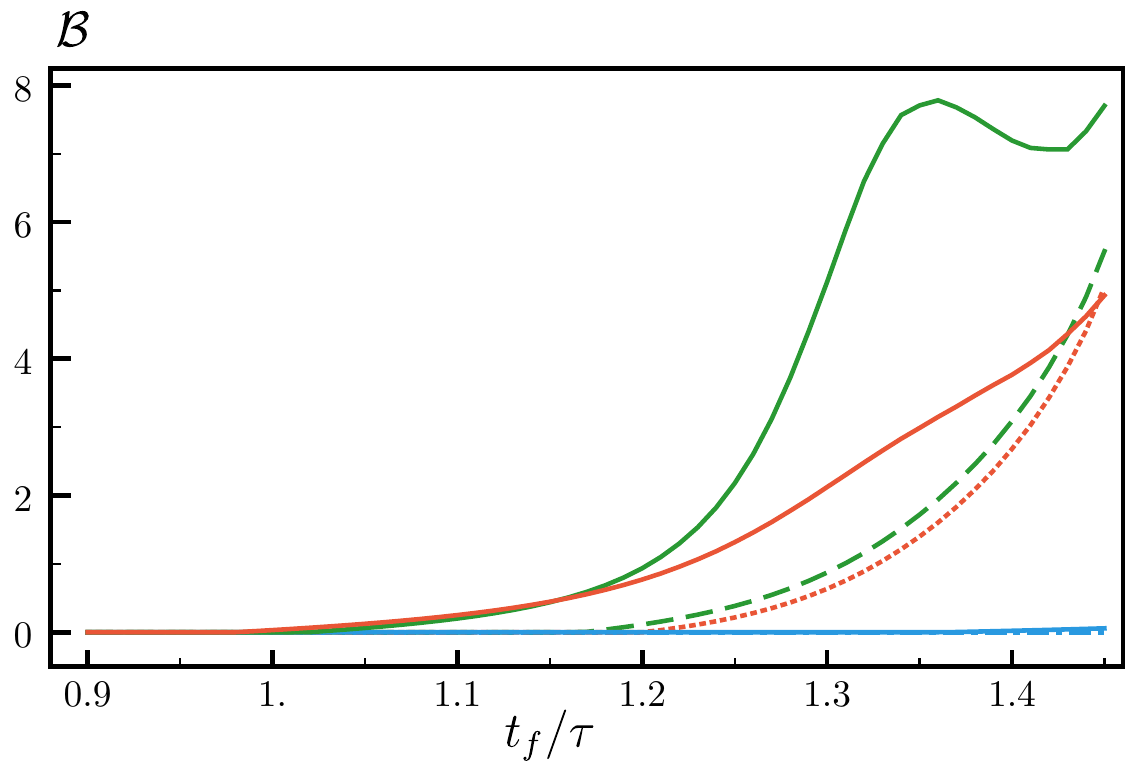}
\end{center}
\caption{\label{fig_7} 
Systematic error bound $\cB$ versus $t_f$ with $F_R=0.9$ for (a) amplitude error $V_{\text{err}}^A$, (b) wavenumber error $V_{\text{err}}^k$, (c) correlated error $V_{\text{err}}^c$.
eSTA trajectories: $Q_2$ (solid green) and $Q_3$ (solid red). STA trajectories: $q_{0,2}$ (dashed green) and $q_{0,3}$ (dotted red).
}
\end{figure}

For a given final time, we approximate a bound $\cB$ on the strength of the systematic error $\delta$, such that for $\fabs{\delta} < \cB$ we ensure that the fidelity satisfies $F > F_R$, with $F_R$ a given reference fidelity.
We can approximate $\cB$ by assuming a linear dependence of the fidelity on $\delta$, with
\begin{align}\label{eq:B}
\cB &= \begin{cases}
\frac{ F(\delta=0) - F_R}{S} &: F(\delta=0) > F_R \\
0                           &: F(\delta=0) \le F_R
\end{cases},
\end{align}
where for simplicity we have used the convention that $\cB = 0$ for $F(\delta=0) \le F_R$.
The systematic error bound $\cB$ indicates that for a given final time, the trajectory achieves fidelity above $F_R$ for $\fabs{\delta}<\cB$.
Hence, higher values of $\cB$ mean higher fidelity and lower sensitivity.
(In contrast, higher values of $S$ correspond with decreased stability at a given $\delta$.)

In Fig \ref{fig_5} we show a specific example of the quantities needed to calculate $\cB$.
For $Q_2$ (solid-green line), the horizontal solid-black arrows in Fig. \ref{fig_5} show $\fabs{\delta}$ symmetric about $\delta=0$ such that $F(\delta)> F_R=0.9$.
To calculate $\cB$, we find the difference $F(\delta=0)-F_R$ (the vertical solid arrow in Fig \ref{fig_5}), and then scale by $1/S$.
Note that for this example, $S$ can be seen on Fig. \ref{fig_6} (c).

The systematic error bound $\cB$ is shown in Fig. \ref{fig_7} for the same trajectories and errors as shown previously in Fig. \ref{fig_6}.
In Fig. \ref{fig_7} we choose $F_R=0.9$ and note that the quasi-optimal eSTA trajectories $Q_2$ and $Q_3$ show significant improvement over the polynomial ansatz eSTA $Q_1$.
We include the polynomial ansatz trajectories $Q_1$ and $q_{0,1}$, as a reference case to highlight the usefulness of $\cB$ as a robustness measure.
For all three errors the eSTA trajectories achieve a higher or equal $\cB$ over their STA counterparts, and this reflects the previous fidelity results (Fig. \ref{fig_3_fid}) and sensitivity results (Fig. \ref{fig_6}).

For $t_f\rightarrow \infty$, for both the eSTA and STA error bounds $\cB \rightarrow \infty$, since as discussed in the last section $S\to 0$ in the adiabatic limit.
This is not an inherent problem with $\cB$, since we are interested in applying $\cB$ in regions far from adiabaticity.
Thus we restrict our investigation to $t_f/\tau<1.45$, as shown in Fig. \ref{fig_7}. 

In Fig. \ref{fig_7} we note that the three error types produce similar $\cB$ behavior for each trajectory, although the magnitude of $\cB$ is different in each case.
We find that from $t_f\ge 1$ the eSTA trajectories $Q_2$ and $Q_3$ give larger values of $\cB$ than the STA trajectories $q_{0,2}$ and $q_{0,3}$.
These values are increasing, indicating that even when the fidelities are very high, a larger value of $\cB$ informs us that the eSTA sensitivity must be decreasing faster than the STA sensitivity.
In this way the error bound $\cB$ gives us a useful comparison between trajectories, as it allows both the fidelity and sensitivity of different trajectories to be compared simultaneously.

\section{Robustness of eSTA to noise \label{sect_robustness_noise}}

We now consider the robustness and stability of eSTA with respect to noise.
Specifically we consider lattice transport with classical Gaussian white noise in position and lattice amplitude.
The noise sensitivity of lattice transport using STA has previously been studied in \cite{luFastShuttlingTrapped2014,luFastShuttlingParticle2018,luNoiseSensitivitiesAtom2020}.

We consider a Hamiltonian $\cH=H_0(t) +\eta \: \xi(t) H_1(t)$, where $\eta$ is the noise strength, $\xi(t)$ is a realization of the noise and $H_1(t)$ is the operator coupling the system to the noise \cite{daviesQuantumStochasticProcesses1969,daviesQuantumStochasticProcesses1970,kielyExactClassicalNoise2021a}.
As shown in Appendix \ref{app:S_N} a master equation can be derived \cite{luFastShuttlingTrapped2014,luFastShuttlingParticle2018}
\begin{align}
\frac{d}{d t} \rho = -\frac{i}{\hbar} \left[ H_0, \rho \right] -\frac{\eta^2}{2\hbar^2} \left[ H_1, \left[H_1, \rho \right] \right],
\end{align}
where $H_0$ is from \Eqref{eq:H_s}.

We define the noise sensitivity $S_N$,
\begin{align}\label{eq:S_N_simple}
S_N = \fabs{\frac{\partial F}{\partial \left( \eta^2 \right)}}
=
\fabs{\frac{1}{2} \frac{\partial^2 F}{\partial \eta^2}},
\end{align}
and in Appendix \ref{app:S_N} we use a perturbation approach to the master equation \cite{yuNonMarkovianQuantumstateDiffusion1999,luFastShuttlingTrapped2014} to obtain
\begin{align}\label{eq:S_N}
S_N &= \frac{1}{\hbar^2} 
\Bigg\vert
\int_0^t ds 
\Big[
\nonumber \\
&\text{Re} \left\{ 
\braXket{\Psi_T(s)}{H_1^2(s)}{\Psi_0(s)}\braket{\Psi_0(s)}{\Psi_T(s)}
\right\}
\nonumber \\
&-
\fabsq{\braXket{\Psi_T(s)}{H_1(s)}{\Psi_0(s)}}
\Big]
\Bigg\vert,
\end{align}
where we use the same notation as in \Eqref{eq:S_systematic_calc} regarding $\ket{\Psi_0(s)}$ and $\ket{\Psi_T(s)}$.
We focus on transport of the ground state of the lattice, but the results in this section can be generalized naturally.
The quantity $S_N$ in \Eqref{eq:S_N} is useful as it allows us measure the system's sensitivity to noise without having to numerically simulate the full open system dynamics, for example using quantum trajectories.

Let us first consider the special case of the adiabatic limit.
Let $\psi_0(x)$ be the ground state of the lattice, thus 
\begin{align}
\Psi_0(t,x) = \Psi_T(t,x) = \psi_0[x-Q_j(t)] e^{i \phi(t)}.
\end{align}
Then \Eqref{eq:S_N} simplifies to
\begin{align}
S_N =t_f C,
\end{align}
where $C$ is given by
\begin{align}\label{eq:S_N_c}
C = \frac{1}{\hbar^2} 
\Bigg\vert
\braXket{\psi_0}{H_1^2}{\psi_0}
-
\fabsq{\braXket{\psi_0}{H_1}{\psi_0}}
\Bigg\vert.
\end{align}

We again consider the error bound defined in the previous section which combines the fidelity and the sensitivity, defined as 
\begin{align}\label{eq:B_N}
\cB_N &= \begin{cases}
\frac{ F(\delta=0) - F_R}{S_N} &: F(\delta=0) > F_R \\
0                           &: F(\delta=0) \le F_R
\end{cases},
\end{align}
where we again adopt the convention $\cB_N = 0$ for $F \le F_R$.
We note that in the adiabatic limit, $\cB_N \approx \frac{1-F_R}{C} \cdot \frac{1}{T}$, i.e. the error bound goes to zero for $T \to \infty$, where $T=t_f/\tau$ (in contrast with the systematic errors considered in the previous section).

\subsection{Position noise}
\begin{figure}[h!]
\begin{center}
(a) \includegraphics[width=\imageScale\columnwidth]{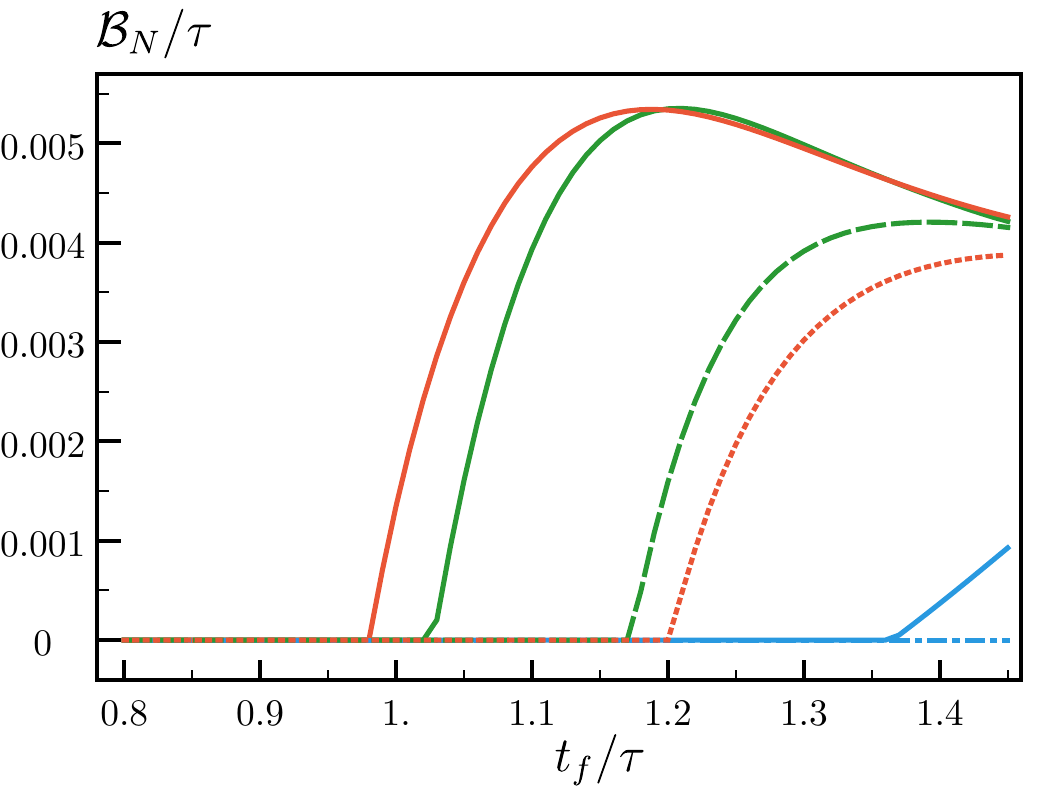}\\
(b) \includegraphics[width=0.4\columnwidth]{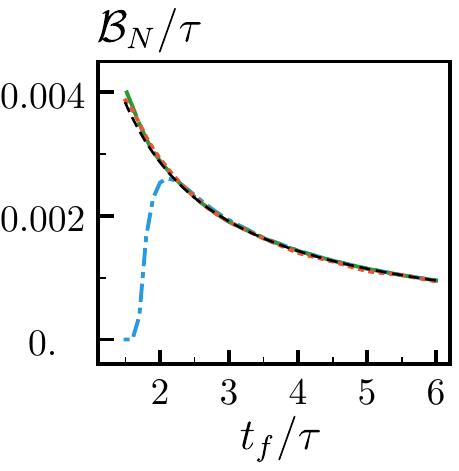}
(c) \includegraphics[width=0.4\columnwidth]{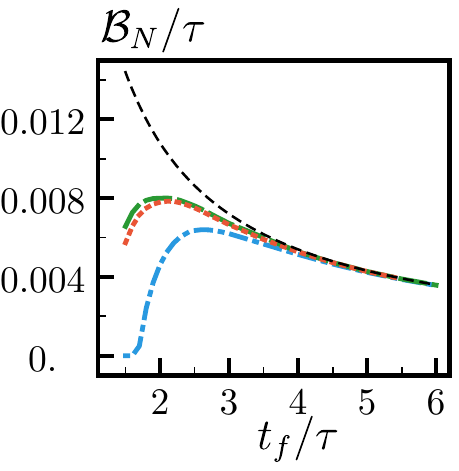}\\
(d) \includegraphics[width=\imageScale\linewidth]{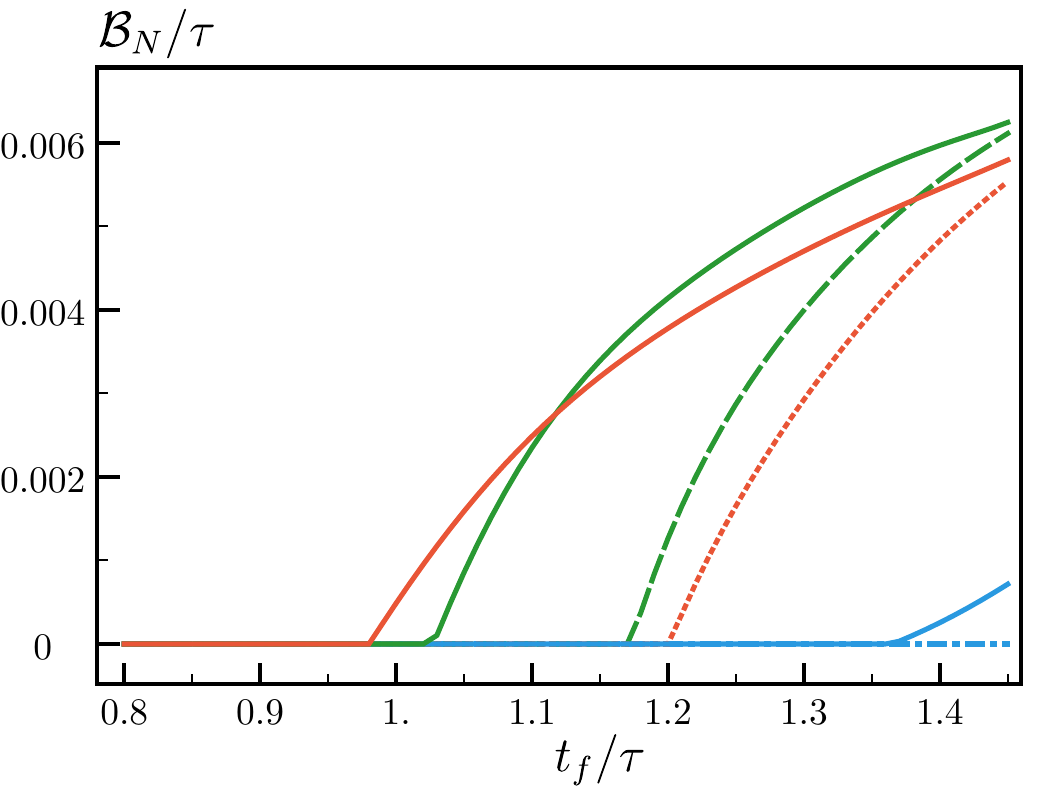}
\end{center}
\caption{\label{fig_8} 
Noise error bound $\cB_N$ versus $t_f$ with $F_R=0.9$ for:
(a) Position noise $V_{N}^P$.
eSTA trap trajectories $Q_{1}(t)$ (solid blue), $Q_{2}(t)$ (solid green) and $Q_{3}(t)$ (solid red);
STA trap trajectories $q_{0,1}(t)$ (dot-dashed blue), $q_{0,2}(t)$ (dashed green) and $q_{0,3}(t)$ (dotted red).\\
(b) Position noise $\cB_N$ versus larger $t_f$ using the STA trajectories. All the trajectories converge to the adiabatic limit (dashed black) for large $t_f$.
(c) Amplitude noise $\cB_N$ versus larger $t_f$, again using the STA trajectories. They also converge to the adiabatic limit (dashed black) for large $t_f$.
(d) Amplitude noise $V_{N}^A$, with same details as (a).
}
\end{figure}

As a first example we consider position noise described by the potential
\begin{align}
V_{N}^P = V[x-Q_j(t)- \sigma \eta \xi (t)],
\end{align}
with $V$ the lattice potential and $\sigma$ the unit of space defined in Sec. \ref{sect_physical_setting}.
Using only first-order in $\eta$, we have
\begin{align}
H_1^P&=-\sigma\frac{\partial}{\partial x} V[x-Q_j(t)]
\nonumber \\
&=
-\sigma U_0 k_0 \operatorname{Sin} \left\{ 2 k_0 [x - Q_j(t)] \right\}.
\end{align}
We evaluate \Eqref{eq:S_N} using both STA and eSTA trajectories and plot $\cB_N$ in Fig. \ref{fig_8} (a).
The eSTA trajectories again out-perform their STA counterparts, showing a greater $\cB_N$ over a larger range of shorter final times.

In Fig. \ref{fig_8} (b) we look at larger $t_f/\tau$ and we see the STA trajectories (dashed-colored lines) approach the adiabatic limit (dashed-black line).

Note that when the ground state is known analytically as with the harmonic oscillator, then explicit formulas for the constant $C$ in \Eqref{eq:S_N_c} can be found.
Using the harmonic oscillator ground state as an approximation to the lattice ground state, we obtain an approximation for \Eqref{eq:S_N_c}
\begin{align}\label{eq:C^P_approx}
C^P &\approx \frac{1}{2}\left( \frac{U_0 k_0 \sigma}{\hbar} \right)^2 \left[ 1 - e^{- 4k_0^2\sigma^2} \right],
\nonumber \\
&=
\frac{\omega_0^2 \widetilde{U}_0}{4}\left[ 1 - e^{- 2/\widetilde{U}_0 } \right],
\end{align}
where $\widetilde{U}_0=U_0/(\hbar \omega_0)$.
For the values considered here, $\tau^2 C_P  = 0.0112$ and the approximation in \Eqref{eq:C^P_approx} gives $\tau^2 C_P  \approx 0.0108$.

\subsection{Amplitude noise}
For noise in the lattice amplitude we define the noise potential as
\begin{align}
V_{N}^A = [1 + \eta \xi (t)] V[x-Q_j(t)],
\end{align}
with $V$ the lattice potential.
In this case we have
\begin{align}
H_{1}^A &= V[x-Q_j(t)].
\end{align}
The results for the STA and eSTA trajectories are shown Fig. \ref{fig_8} (d).
As with the position noise, the eSTA trajectories are an improvement over the STA trajectories.

Both noise sources have similar $\cB_N$ scales, with the position noise eSTA results showing greater improvement over STA than the amplitude noise eSTA results have over their STA counterparts

In Fig. \ref{fig_8} (c) the STA trajectories for amplitude noise are shown for values of $t_f/\tau$ approaching the adiabatic limit, and for $t_f/\tau >4$ they agree.

As with the previous section, we can approximate $C^A$ by using the analytic harmonic ground state with the lattice potential,
\begin{align}
C^A &\approx 
\frac{1}{8}
\left(\frac{U_0}{\hbar}\right)^2
e^{-4 k_0^2 \sigma^2} 
\left(
e^{2 k_0^2 \sigma^2} - 1
\right)^2,
\nonumber \\
& =
\frac{\omega_0^2 \widetilde{U}_0^2}{8}
e^{-2 / \widetilde{U}_0} 
\left(
e^{1/\widetilde{U}_0} - 1
\right)^2,
\end{align}
and again $\widetilde{U}_0=U_0/(\hbar \omega_0)$.
The approximation gives $\tau^2 C^A\approx 2.70\times 10^{-3}$, while the exact value is $\tau^2 C^A=2.97\times 10^{-3}$.

\section{Conclusion}

We applied the general eSTA formalism developed in \cite{whittyQuantumControlEnhanced2020} to the practical problem of atom transport using an optical lattice potential, near the quantum speed limit \cite{peterDemonstrationQuantumBrachistochrones2021}.
By examining the robustness of the eSTA control schemes, we found that the eSTA transport protocols result in higher fidelity and improved robustness against several types of systematic errors and noise errors.

We have provided a general heuristic argument that the eSTA schemes should result in higher fidelities and improved stability compared with the original STA schemes.
Furthermore, we have shown strong numerical evidence of this claim by considering noise and systematic errors in the lattice potential.

Finally, we have quantified this increased robustness by defining new measures. These include an eSTA- specific evaluation tool
$C_Q$, that allows possible control functions to be evaluated without full numerical treatment and a practical error bound $\cB$ that combines fidelity and sensitivity such that eSTA and STA control functions can be compared qualitatively
In the future the robustness of eSTA schemes and the error bound $\cB$ could be considered in further quantum control applications.

\begin{acknowledgments}
We are grateful to D. Rea for useful discussion and commenting on the manuscript.
C.W. acknowledges support from the Irish Research Council (GOIPG/2017/1846) and support from the Google Cloud Platform research credits program.
A.K. acknowledges support from the Science Foundation Ireland Starting Investigator Research Grant ``SpeedDemon'' No. 18/SIRG/5508.
A.R. acknowledges support from the Science Foundation Ireland Frontiers for the Future Research Grant ``Shortcut-Enhanced Quantum Thermodynamics'' No. 19/FFP/6951.
\end{acknowledgments}

\begin{appendix}

\begin{widetext}
\section{Derivation of $\cC_Q$}\label{app:s_Q}
We can evaluate $\cC_Q$ using the formalism of eSTA.
As before with eSTA, we write the system Hamiltonian $\cH(\delta)$ as an expansion about the STA system that approximates it
\begin{align}
\cH(\delta) = H_0^{\text{STA}}(\delta) + \mu \cH_1(\delta)  + \mu^2 \cH_2(\delta)  +\dots
\end{align}
We generalize the definitions of $\vec{K}_n$ and $G_n$ from the eSTA formalism to include the $\delta$ dependence in the Hamiltonian, giving
\begin{eqnarray}
\vec{K}_n \left( \delta \right) &=& \int_0^{t_f} dt \, \bra{\chi_n (t)} \nabla \Hamil_S (\vec\lambda_0; t; \delta) \ket{\chi_0 (t)},
\label{eq_Kn}
\end{eqnarray}
and
\begin{eqnarray}
G_n \left( \delta \right) = \!\! \int_0^{t_f} \!\! dt \braXket{\chi_n(t)}{\left[\Hamil_S (\vec\lambda_0;t;\delta) - \Hamil^{(0)} (\vec\lambda_0;t)\right]}{\chi_0(t)}.
\label{eq_Gn}
\end{eqnarray}
Hence we have that the eSTA control vector is given by
\begin{eqnarray}
\vec\epsilon \left( \delta \right) 
\approx 
-
\frac{
\left[
\sum_{n=1}^N \fabsq{G_n \left( \delta \right)}
\right]
\left\{
\sum_{n=1}^N \mbox{Re} \left[ G_n^* \left( \delta \right)\vec{K}_n\left( \delta \right) \right]
\right\}
}{
\fabsq{\sum_{n=1}^N \mbox{Re} \left[G_n^* \left( \delta \right)\vec{K}_n\left( \delta \right)\right]}
}.
\label{epsilon_del}
\end{eqnarray}
Now we look at the derivative with respect to $\delta$ and evaluate at $\delta=0$,
\begin{eqnarray}
\frac{\partial}{\partial \delta} \epsilon_{j}\left(0\right) 
=\epsilon_{j}\left(0\right) 
\Bigg(
\frac{
\sum_{n=1}^N 2\mbox{Re} 
\left[
G_n^* \left( 0 \right) \frac{\partial}{\partial \delta}G_n\left( 0 \right)
\right]
}{
\sum_{n=1}^N \fabsq{
G_n \left( 0 \right)
}
}
+
\frac{
\sum_{n=1}^N 
\left\{
\mbox{Re} 
\left[
K_{n,j}^* \left( 0 \right) \frac{\partial}{\partial \delta}G_n\left( 0 \right)
\right]
+
\mbox{Re}
\left[
G_n^* \left( 0 \right) \frac{\partial}{\partial \delta}K_{n,j}\left( 0 \right)
\right]
\right\}
}{
\sum_{n=1}^N \mbox{Re} 
\left[
G_n^* \left( 0 \right)\vec{K}_{n,j}\left( 0 \right)
\right]
}
\nonumber \\ 
+
\frac{
\sum_{j=1}^\text{M}
\left\{
\sum_{n=1}^N \mbox{Re} 
\left[
G_n^* \left( 0 \right)\vec{K}_{n,j}\left( 0 \right)
\right]
\sum_{n=1}^N 
\mbox{Re} 
\left[
K_{n,j}^* \left( 0 \right) \frac{\partial}{\partial \delta}G_n\left( 0 \right)
\right]
+
\mbox{Re} 
\left[
G_n^* \left( 0 \right) \frac{\partial}{\partial \delta}K_{n,j}\left( 0 \right)
\right]
\right\}
}{
\sum_{j=1}^M
\fabsq{\sum_{n=1}^N \mbox{Re} 
\left[
G_n^* \left( 0 \right)\vec{K}_{n,j}\left( 0\right)
\right]}.
}
\Bigg)
\end{eqnarray}
While this expression is detailed, the individual terms are known exactly and can be calculated entirely analytically.
\end{widetext}

\section{Derivation of $S_N$}\label{app:S_N}
To consider the effect of noise on eSTA protocols, we start with a system obeying the Schr\"odinger equation
\begin{align}
i \hbar \frac{\partial}{\partial t} \ket{\Psi} = H_0(t) \ket{\Psi},
\end{align}
where $H_0(t) = p^2/2m + V \left(x,t \right)$, and $V$ is the potential.
We consider noise of the form $\eta \: \xi(t) H_1(t)$, where $\xi(t)$ is a given noise realization, $H_1(t)$ is the operator coupling the system to the noise and $\eta$ is the noise strength \cite{daviesQuantumStochasticProcesses1969,daviesQuantumStochasticProcesses1970,luFastShuttlingTrapped2014}.
In this paper we assume that the statistical expectation $\cE [\xi(t) ] = 0$, with
\begin{align}
\cE [ \xi(t) \xi(t') ] = \alpha(t - t'),
\end{align}
where $\alpha(t - t')$ is the correlation function of the noise. Following the approach taken in \cite{yuNonMarkovianQuantumstateDiffusion1999}, a master equation can be derived \cite{luFastShuttlingTrapped2014,luFastShuttlingParticle2018}
\begin{align}
\frac{d}{d t} \rho = -\frac{i}{\hbar} \left[ H_0, \rho \right] -\frac{i \eta}{\hbar} \left[ H_1, \langle\xi \rho\rangle \right],
\end{align}
where $\eta$ is the perturbation parameter and $\rho$ is the average over realizations of $\xi(t)$.
Now we assume Gaussian white noise, i.e. $\alpha(t - t')=\delta(t'-t)$ and using Novikov's theorem \cite{novikov}, $\langle \xi \rho\rangle  = -i \eta / 2 \hbar [H_1,\rho]$ we obtain
\begin{align}
\frac{d}{d t} \rho = -\frac{i}{\hbar} \left[ H_0, \rho \right] -\frac{\eta^2}{2\hbar^2} \left[ H_1, \left[H_1, \rho \right] \right].
\end{align}
We define
\begin{align}
\frac{d}{d t} \rho_0 = - \frac{i}{\hbar} \left[ H_0, \rho_0 \right],
\nonumber \\
\Sket{\rho_0(t)} = U_0(t,0) \Sket{\rho_0(0)},
\end{align}
where $\Sket{\rho_0(t)}$ denotes $\rho_0(t)$ written in super-operator notation.
Let
\begin{align}\label{eq:L_noise}
\cL(t)\Sket{\rho} = -\frac{1}{2\hbar^2} \left[ H_1, \left[H_1, \rho \right] \right],
\end{align}
then \cite{yuNonMarkovianQuantumstateDiffusion1999} \cite{luFastShuttlingTrapped2014}
\begin{align}\label{eq:rho}
\Sket{\rho(t)} &= \Sket{\rho_0(t)} + \eta^2 \int_0^t ds U_0(t,s) \cL(s) U_0(s,0) \Sket{\rho_0(0)}
\nonumber \\
&+ \cO(\eta^4)
\nonumber \\
&= \Sket{\rho_0(t)} + \eta^2 \int_0^t ds U_0(t,s) \cL(s) \Sket{\rho_0(s)} + \cO(\eta^4).
\end{align}
We denote the target state as $\ket{\Psi_T}$ and set $\Sket{\rho_T} = \ketbra{\Psi_T}{\Psi_T}$.
The fidelity is then
\begin{align}
\fid &= \Sbraket{\rho_T}{\rho}=\text{Tr}\left( \rho_T^\dagger \rho\right) = \braXket{\psi_T}{\rho(t)}{\psi_T}
\nonumber \\
&= \Sbraket{\rho_T}{\rho_0} + \eta^2 \int_0^t ds \Sbra{\rho_T(s)} \cL(s) \Sket{\rho_0(s)} 
\nonumber \\
&+ \cO(\eta^4).
\end{align}
We define the noise sensitivity $S_N$,
\begin{align}
S_N = \left\vert \frac{\partial \fid}{\partial \left( \eta^2 \right)} \right\vert = \left\vert \int_0^t ds \Sbra{\rho_T(s)} \cL(s) \Sket{\rho_0(s)}  \right\vert.
\end{align}
Using $\rho_0=\ketbra{\Psi_0}{\Psi_0}$ and the explicit form of $\cL$ from \Eqref{eq:L_noise}, this expression can be simplified to
\begin{align}
S_N &= \frac{1}{\hbar^2} 
\Bigg\vert
\int_0^t ds 
\Big[
\nonumber \\
&\text{Re} \left\{ 
\braXket{\Psi_T(s)}{H_1^2(s)}{\Psi_0(s)}\braket{\Psi_0(s)}{\Psi_T(s)}
\right\}
\nonumber \\
&-
\fabsq{\braXket{\Psi_T(s)}{H_1(s)}{\Psi_0(s)}}
\Big]
\Bigg\vert.
\end{align}

\end{appendix}

\twocolumngrid

\bibliography{eSTA_robustness}{}
\bibliographystyle{apsrev4-1}

\end{document}